\documentclass[]{aa}
\usepackage{graphicx}
\usepackage{natbib}
\usepackage[fleqn]{amsmath}
\usepackage[modulo, switch]{lineno}
\usepackage{nicefrac}
\usepackage{textcomp}
\usepackage{xfrac}

\usepackage{txfonts}
\def\mh{\,$\mu$Hz}
\def\ph{\,ppm$^2$/$\mu$Hz}
\def\teff{$T_{\mathrm{eff}}$}
\def\lg{\ensuremath{\log g}}
\def\num{$\nu_\mathrm{max}$}

\def\sun{\hbox{$_\odot$}}

\begin{document}
%
   \title{The connection between stellar granulation and oscillation as seen by the \textit{Kepler} mission}

   \author{T. Kallinger\inst{1}
 	  \and
	  J. De Ridder\inst{2}
	  \and
	  S. Hekker\inst{3,4}
	  \and
	  S. Mathur\inst{5,6}
	  \and
	  B. Mosser\inst{7}
	  \and
          M. Gruberbauer\inst{8, 1}
          \and
          R. A. Garc\'ia\inst{9}
          \and
	  C. Karoff\inst{10,11}
	\and 
	J. Ballot\inst{12,13}
            }

   \offprints{thomas.kallinger@univie.ac.at}

   \institute{
Institute for Astronomy (IfA), University of Vienna, T\"urkenschanzstrasse 17, 1180 Vienna, Austria
              	\and
Instituut voor Sterrenkunde, KU Leuven, Celestijnenlaan 200D, 3001 Leuven, Belgium
		\and
Max-Planck-Institut f\"ur Sonnensystemforschung, Justus-von-Liebig-Weg 3, 37077 G\"ottingen, Germany
		\and
Astronomical Institute ``Anton Pannekoek", University of Amsterdam, PO Box 94249, 1090 GE Amsterdam, The Netherlands
		\and
High Altitude Observatory, NCAR, P.O. Box 3000, Boulder, CO 80307, USA
		\and
Space Science Institute, 4750 Walnut street Suite 205, Boulder, CO 80301, USA
		\and
LESIA, CNRS, Universit\'e Pierre et Marie Curie, Universit\'e Denis Diderot, Observatoire de Paris, 92195 Meudon cedex, France
		\and
Department of Astronomy and Physics, Saint Marys University, Halifax, NS B3H 3C3, Canada
	\and
Laboratoire AIM, CEA/DSM-CNRS, Universit\'e Paris 7 Diderot, IRFU/SAp, Centre de Saclay, 91191, Gif-sur-Yvette, France
	\and
Stellar Astrophysics Centre, Department of Physics and Astronomy, {\AA}rhus University, Ny Munkegade 120, DK-8000 {\AA}rhus C, Denmark
	\and
Department of Geoscience, {\AA}rhus University, H{\o}egh-Guldbergs Gade 2, 8000 {\AA}rhus C, Denmark
	\and
CNRS, Institut de Recherche en Astrophysique et Plan\'etologie, 14 avenue \'Edouard Belin, 31400 Toulouse, France	
	\and
Universit\'e de Toulouse, UPS-OMP, IRAP, Toulouse, France
}

   \date{Received  30 May 2014 / Accepted 31 July 2014}

\abstract
{The long and almost continuous observations by \textit{Kepler} show clear evidence of a granulation background signal in a large sample of stars, which is interpreted as the surface manifestation of convection. It has been shown that its characteristic timescale and rms intensity fluctuation scale with the peak frequency (\num ) of the solar-like oscillations. Various attempts have been made to quantify the observed signal, to determine scaling relations for its characteristic parameters, and to compare them to theoretical predictions.}
{We aim to study different approaches to quantifying the signature of stellar granulation and to search for a unified model that reproduces the observed signal best in a wide variety of stars. We then aim to define empirical scaling relations between the granulation properties and \num\ and various other stellar parameters.}
{We use a probabilistic method to compare different approaches to extracting the granulation signal. We fit the power density spectra of a large set of \textit{Kepler} targets, determine the granulation and global oscillation parameter, and quantify scaling relations between them.}
{We establish that a depression in power at about \num /2, known from the Sun and a few other main-sequence stars, is also statistically significant in red giants and that a super-Lorentzian function with two components is best suited to reproducing the granulation signal in the broader vicinity of the pulsation power excess. We also establish that the specific choice of the background model can affect the determination of \num , introducing systematic uncertainties that can significantly exceed the random uncertainties. We find the characteristic frequency and amplitude of both background components to tightly scale with \num\ for a wide variety of stars, and quantify a mass dependency of the latter. To enable comparison with theoretical predictions, we computed effective timescales and bolometric intensity fluctuations and found them to approximately scale as $\tau_\mathrm{eff} \propto g^{-0.85}\,T^{-0.4}$ and $A_\mathrm{gran} \propto (g^2M)^{-1/4}$ (or more conveniently $R/M^{3/4}$), respectively. Similarly, the bolometric pulsation amplitude scales approximately as $A_\mathrm{puls} \propto (g^2M)^{-1/3}$ (or $R^{4/3}/M$), which implicitly verifies a separate mass and luminosity dependence of $A_\mathrm{puls}$.}
{We provide a thorough analysis of the granulation background signal in a large sample of stars, from which we establish a unified model that allows us to accurately extract the granulation and global oscillation parameter.}

\authorrunning{Kallinger et al.}
\titlerunning{The connection between stellar granulation and oscillations}
   \maketitle

\section{Introduction}	\label{sec:intro}

Modern theoretical physics is still lacking a complete description of turbulence, which has proven to be highly resistant to full modelling because of the wide range in scale from macroscopic to microscopic. A promising opportunity for studying this process is provided by stellar astrophysics. Turbulent convection is an important energy transport mechanism that is essential for modelling the structure and evolution of our Sun and stars in general.

The surface of the Sun shows an irregular cellular pattern, which is known as the surface-visible signature of convection. It is due to hot plasma that rises from the outer convective zone to the photosphere, where it forms bright cells (the so-called granules), cools down, and descends at the darker inter-granular lanes. These granulation structures on the solar surface were first discovered by \citet{herschel}, and thanks to numerical hydrodynamical simulations, their properties are nowadays explained well \citep[e.g.,][]{muller99}. Other phenomena related to convection are the so-called solar-like oscillations that are due to turbulent motions in the convective envelope producing an acoustic noise that stochastically drives resonant, intrinsically damped acoustic oscillations \citep[e.g.,][]{jcd02}. 

In the Hertzsprung-Russel (HR) diagram, the red border of the classical $\delta$ Sct instability strip (IS$_\mathrm{RB}$) is expected to mark the transition from opacity mechanisms driving stellar oscillations on the blue side of the IS$_\mathrm{RB}$ to stochastic excitation on the red side. This is also the region where stars develop an appreciable upper convective envelope giving rise to surface granulation. In contrast to this simplistic picture, evidence has been found that even the very shallow convective envelopes of hot stars ($<$1\% in radius) are sufficient to produce an observable granulation signal \citep{kal10d} and potentially excite solar-like oscillations \citep{antoci11,bel09}. 

However, all stars with an convective envelope show granulation. The granules on their surface evolve with time and produce quasi-periodic brightness fluctuations on a wide range of timescales and amplitudes \citep[see e.g.,][for the Sun]{har85,aigrain04}. Thanks to the high precision photometric measurements of CoRoT \citep{bag06} and \textit{Kepler} \citep{bor10,koch10}  such brightness fluctuations are subject to continuous long-term monitoring for a large variety of stars. First results for a large sample study have been presented by \citet{mat11}, who investigated the characteristics of granulation in red giants observed by \textit{Kepler}. They found that the characteristic timescale ($\tau_\mathrm{eff}$) and the rms brightness fluctuation ($\sigma$) of the granulation signal scale to a first approximation with the peak frequency (\num) of the solar-like oscillations\footnote{The peak frequency (\num ; also known as the frequency of maximum oscillation power) is expected to be proportional to the acoustic cutoff-frequency ($\nu_{ac} \propto g/\sqrt{T_\mathrm{eff}}$) of the stellar atmosphere \citep[e.g.,][]{bro91,kje95,bel11}.} as $\tau_\mathrm{eff} \propto \nu_\mathrm{max}^{-0.89}$ and $\sigma \propto \nu_\mathrm{max}^{-0.45}$, respectively, which is consistent with basic theoretical predictions \citep[e.g.,][]{kjeldsen11,cha11}. The authors also performed hydrodynamical 3D simulations of granulation in red giants on the basis of the \citet{ludwig2006} \textit{ab initio} approach. Even though these simulations match the observed scaling relations well in terms of trends, they found large systematic differences of up to a factor of three. Recently, \citet{sam13a,sam13} provided a simple theoretical model that supports the observed variations of the granulation properties with \num , but also suggests that the turbulent Mach number in the photosphere plays an important role in controlling granulation. As in \citet{mat11}, they found their theoretical granulation parameters to match the observations well on a global scale but to systematically underestimate the measurements. From these attempts it appears that the general trends of the physical properties of granulation across the HR diagram are reproducible from a modelling point of view, but the source of the rather large systematic deviations remains unknown. The reason for that is not necessarily entirely due to insufficient models. There is also no consensus yet on how to exactly extract the granulation parameters from the observations and large systematic differences have been found between different methods \citep[see][for a comparison]{mat11}.

To infer the characteristics of stellar granulation from the power spectrum of a stars' intensity variations is not straightforward and different analysis methods are found in the literature. These methods use one or more components to model the background signal locally around \num , and differ from each other not only in the number of components used but also in the functional form of the individual components. Usually models of the form $P(\nu) \propto 1/[1+(\pi\tau\nu)^c]$ are implemented, in which the exponent $c$ plays an important role as it controls how fast the power, $P$, decays with increasing frequency. Originally \citet{har85} adopted a value of two (i.e., a Lorentzian function, which is in this context also often named the ``classical Harvey model'') to model the solar background signal but already mentioned that the actual exponent may well be different from two. The reason why this is of particular interest is because the exponent defines the general shape of the background component and the use of an ``incorrect'' value can easily result in systematic deviations of the other background parameters. This effect has not yet been investigated thoroughly. In fact, it is difficult to measure the actual exponent from the observations (even for the Sun). Therefore values different from two were established empirically, like a fixed exponent of four \citep[e.g.,][]{mic09,kal10a} but also more general models with $c$ being a free parameter \citep[resulting in values as high as 6.2 for the Sun,][]{karoff12}.

Another problem arises from the number of components that are superposed to reproduce the stellar background signal. Observational evidence for more than one background component in the vicinity of the pulsation power excess is given by a slight depression in power at about \num/2. Such a feature has been first found in the solar irradiance data from the SOHO/VIRGO instrument \citep[e.g.,][]{vaz05} but is meanwhile also found to be significant in other stars \citep[e.g.,][]{karoff13}. The physical origin of this kink in the power spectrum (and therefore an additional signal component) is still a matter of debate and is either attributed to the occurrence of bright points \citep[e.g.,][]{har93}, faculae \citep{karoff12}, changing properties of the granules \citep{and98}, or a second granulation population \citep[e.g.,][]{vaz05}. 

A more phenomenological approach of granulation was recently demonstrated by \citet{bast13}, who filtered the \textit{Kepler} time series with a 8 hour high-pass filter and found that the rms scatter of the residuals are a good indicator for the surface gravity of the stars. The advantage of this method is that it is much faster and easier to apply than asteroseismic techniques and that it gives a more accurate surface gravity than the usual spectroscopic methods.

The aim of the current analysis is twofold. Firstly, we investigate various approaches to model the granulation background signal in a large sample of stars observed by \textit{Kepler}. The sample includes stars from the main sequence to high up the giant branch and asymptotic giant branch \citep[compared to the analysis in][which was restricted to red giants]{mat10} and therefore basically covers the entire parameter space for which we can expect solar-like oscillations and surface granulation. We use a probabilistic approach to search for a unified model that reproduces the observed signal best for all stars in our sample, where the complexity of the model should be driven by the data quality alone. We find that the superposition of two super-Lorentzian functions (i.e, Harvey-like models with $c=4$) is well suited to reproduce the granulation signal in the broader vicinity of the pulsation power excess (about 0.1 to 10 $\times$ \num ) and fit this model to all stars in our sample.

Secondly, we investigate the resulting granulation timescales and rms brightness fluctuations and find tight correlations between these parameters and \num . We then define empirical scaling relations between the granulation properties and the surface gravity (and other fundamental parameters) of a star and compare them to theoretical predictions. Furthermore, we study the pulsation amplitudes and how they scale with other parameters of the star. 

\section{Observations}	\label{sec:obs}

The \textit{Kepler} space telescope was launched in March 2009 with the primary goal of searching for transiting Earth-sized planets in and near the habitable zones of Sun-like stars. \textit{Kepler} houses a 95-cm aperture telescope that points at a single field in the constellations of Cygnus and Lyra, feeding a photometer with a 115 square-degree wide field of view (FOV) to continuously monitor the brightnesses of over 145\,000 stars. The exquisite precision and accuracy of the photometry makes \textit{Kepler} also an ideal instrument for asteroseismology and the Kepler Asteroseismic Science Consortium (KASC) has been set up to study many of the observed stars (see \citealt{gil10} for an overview and first results).

\textit{Kepler} observations are subdivided into quarters, starting with the initial commissioning run (10\,d, Q0), followed by a short first quarter (34\,d, Q1) and subsequent full quarters of 90\,d length. Photometry of most of the stars is conducted at a long cadence (LC) of 29.42 minutes, but a subset of up to 512 stars can be observed at a short cadence (SC) of 58.82 seconds \citep[for more details see, e.g.,][]{jen10}. Our studies are primarily based on the LC data for 1289 red-giant stars spanning from Q0 to Q13 with a total of about 51\,800 measurements per star. Apart from occasional losses of the satellite's fine pointing and scheduled re-orientations of the spacecraft, the about 1142 day-long observations were continuous, with an overall duty cycle of about 93\,\%. The sample includes only stars that show a clear power excess due to pulsation and which have already been used for other studies, such as the investigation of different seismic \citep{hub10,hub11,mos12a} and granulation \citep{mat11} observables, the seismic determination of fundamental parameters \citep{kal10b}, a comparison of global oscillation parameters derived from different methods \citep{hek10b,hek12}, a detailed analysis of an individual star \citep{diMauro11} and of the radial \citep{kal12} and non-radial mixed mode spectra \citep{bed11,mos12b}, and how the latter are used to constrain rotational properties \citep{beck11b} and their evolution \citep{mos12c}.

\begin{figure}[t]
	\begin{center}
	\includegraphics[width=0.5\textwidth]{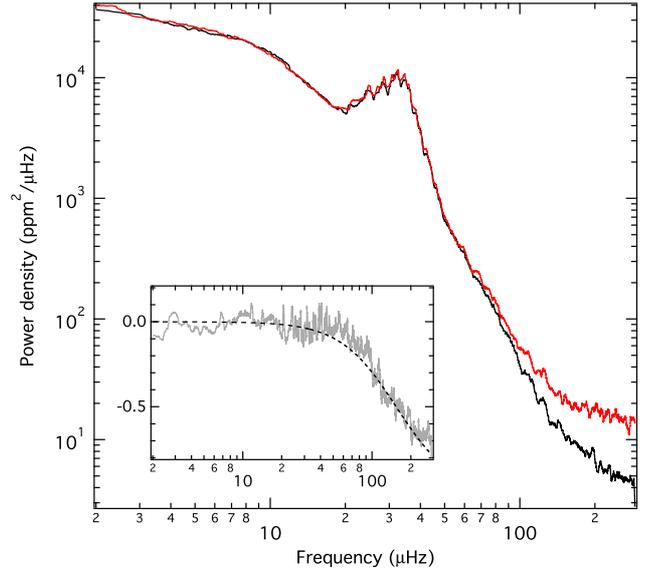}
	\caption{Comparison between the heavily smoothed power density spectra of KIC 5517118 based on ``noise-optimised'' (black line) and ``stability-optimised'' (red line) data, showing additional frequency-dependent (instrumental) noise in the latter. The insert shows the relative difference (PDS$_\mathrm{noise}$-PDS$_\mathrm{stab.}$)/PDS$_\mathrm{stab.}$ and a Lorentzian fit (dashed line).} 
	\label{fig:bgQ10-13} 
	\end{center} 
\end{figure}

Naturally, this sample contains only stars that oscillate with timescales longer than about 1\,h (twice the sampling rate) and is therefore limited to stars on the giant branch. To extend the sample towards the subgiant branch and main sequence we also include SC data of a subsample of solar-type and subgiant stars presented by \citet{chap11b}. The data were obtained between Q5 an Q8 and cover up to one year of continuous observations with up to 480\,000 individual measurements per star. Out of the original sample (500 stars), we selected those for which we have at least one full quarter of observations. Out of the remaining 113 stars we selected the ``best'' 75 stars, that show a clear power excess with reasonably high signal-to-noise ratios and more importantly also the high-frequency part (see Fig.\,\ref{fig:bgQ10-13} above $\sim$50\mh\ ) of the granulation signal (i.e., signal with timescales shorter than the pulsations) is well above the white noise. The latter is quite often not the case, even for the high-precision observations of \textit{Kepler}. This does not result in a selection effect, so that we select only stars with a high granulation amplitude and ignore those with low amplitudes. The fact that we see the high-frequency part of the granulation signal is predominantly due to a low white noise level, which basically reflects the apparent magnitude of the stars. While the original sample roughly covers stars in the range $6.5 \lesssim K_p \lesssim 12$ mag, our SC sample is limited to $K_p \lesssim 10$ mag. 

\section{Data analysis}	\label{sec:data}

\subsection{Flux extraction}

Generally spoken, the \textit{Kepler} raw data are prepared so that the point-to-point scatter of the individual measurements is minimised (``noise-optimised''), which is important to detect and characterize planetary transits. This implies that the apertures on the CCD's from which the photometry is extracted are kept generally small to include only a minimum of the sky background signal. However, this leads to a number of problems. So it happens for example that a target star (but also background stars) drifts across the aperture on the CCD leading to long-term trends in the extracted photometry. Additionally, artefacts are introduced by occasional losses of the satellite fine pointing or during the thermal recovery after \textit{Kepler} is pointed to the Earth. Even though these instrumental effects can be corrected to some extent \citep[see, e.g.,][and references therein]{gar11,smith12}, they prevent us from accessing intrinsic variability timescales longer than a few weeks. This is not a problem for most of the stars in our sample. It, however, limits the sample to stars well below the tip of the giant branch. To improve the situation Mathur et al. (2014, in preperation) developed a new approach, whose basic principle is to adopt the individual apertures so that trends and jumps are minimised (``stability-optimised''). Subsequently, we smooth the data with a triangular filter to suppress residual instrumental long-term trends with timescales longer than about 40\,d. 

The improved long-term stability comes, however, at the price of additional high-frequency noise. This is illustrated in Fig.\,\ref{fig:bgQ10-13}, where we compare the power density spectra of a particular star based on the original ``noise-optimised'' (black line) and the new ``stability-optimised'' (red line) data reductions. Apart from the different apertures the data sets are treated equally. Clearly, the new approach adds significant power at high frequencies\footnote{Note that the improved long-term stability is not noticable in this plot as the differences are well below 2\mh }. The relative difference of the two spectra reveals that the additional noise component is frequency dependent but follows a simple Lorentzian. In the modelling of the power density spectra (see Sec.\,\ref{BGmodel} below) it should therefore be sufficient to add a frequency dependent term to the usually used constant white noise component. 

We note that the presented example is a extreme case. In fact, the additional noise is negligible (i.e., its amplitude is much smaller than the white noise) for most stars in our sample. We also did not find any correlations between the amplitudes or timescales of the Lorentzian and any other fitted parameter and assume that this noise feature is largely defined by how exactly the aperture was chosen and how the sky background looks like in this aperture. However, there is a tendency that the amplitude becomes larger for stars with low \num , which are usually also the brightest stars with a low white noise.

\begin{figure}[t]
	\begin{center}
	\includegraphics[width=0.5\textwidth]{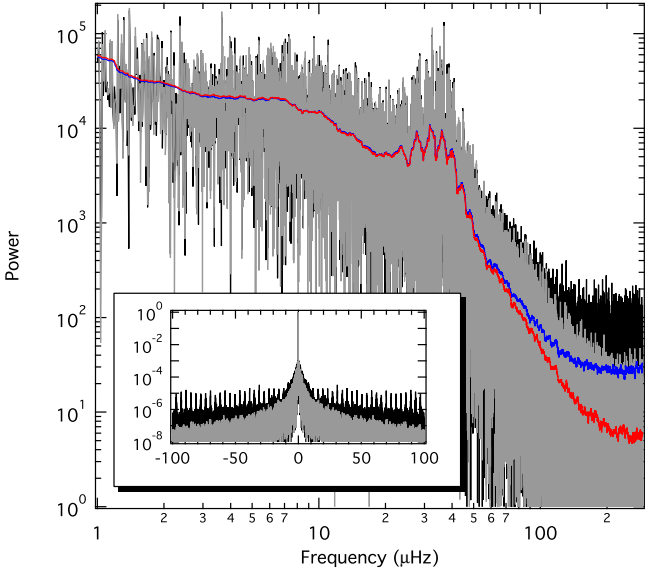}
	\caption{Power spectrum of KIC 7949599. The black and grey lines indicate the spectrum of the raw data (duty cycle of $\sim$93\%) and after filling all gaps shorter than 1 day (duty cycle of $\sim$95\%), respectively, showing how low-frequency signal leaks into the high-frequency range via the spectral window (see insert). The blue and red lines correspond to heavily smoothed versions of the raw and gap-filled spectrum, respectively.} 
	\label{fig:gapfilling} 
	\end{center} 
\end{figure}

\subsection{Gap filling}

Even though \textit{Kepler} is supposed to observe continuously, there are several operations affecting the scientific data acquisition producing quasi-regular gaps \citep[see, e.g.,][]{garcia14}. If we were to keep these gaps the spectral window would be degraded with, e.g., up to 0.5\% high regular peaks (see the insert of Fig.\,\ref{fig:gapfilling}). This might seem negligible but we have to keep in mind that the power contrast between the low-frequency part of the spectrum with its typically high amplitudes and the high-frequency white noise is at least for red giants several orders of magnitudes. Leaving the data gaps untouched would smear out the intrinsic structure in the high-frequency part of the spectrum. To correct for this one can fill the gaps by, e.g., linear interpolation. In Fig.\,\ref{fig:gapfilling} we compare the raw and gap-filled spectrum of a typical \textit{Kepler} data set of a red giant. Clearly, the aliasing signal from the gaps in the original time series buries the structure of the spectrum where the granulation signal fades into the white noise.
On the other hand, filling gaps as long as, e.g., 16 days (resulting from a malfunction of the satellite at the beginning of Q8) would introduce spurious modulation peaks. As a compromise we only fill gaps by linear interpolation if they are shorter than 3/\num\ ($\sim$1 day for most of the red giants), which improves the duty cycle from about 93 to about 95\%. This appears as a small improvement but as can be seen from Fig.\,\ref{fig:gapfilling} it is sufficient to correct for most of the objectionable regularities in the spectral window function. A detailed description of the impact of nominal regular gaps in the \textit{Kepler} time series is given in \citet{garcia14}. We have to mention that gap-filling has only a minor effect on the extraction of the parameters relevant for the subsequent analysis but it becomes important if one wants to study the detailed structure of the granulation background signal (see Sec.\,\ref{Sec:ModelComp}).  

The power spectra are then computed using a Discrete Fourier Transform algorithm \citep[DFT;][]{dem} and converted to power density (see Appendix \ref{PDSconversion}).

\subsection{Sampling effects} \label{AmpDamp}
\begin{figure}[b]
	\begin{center}
	\includegraphics[width=0.5\textwidth]{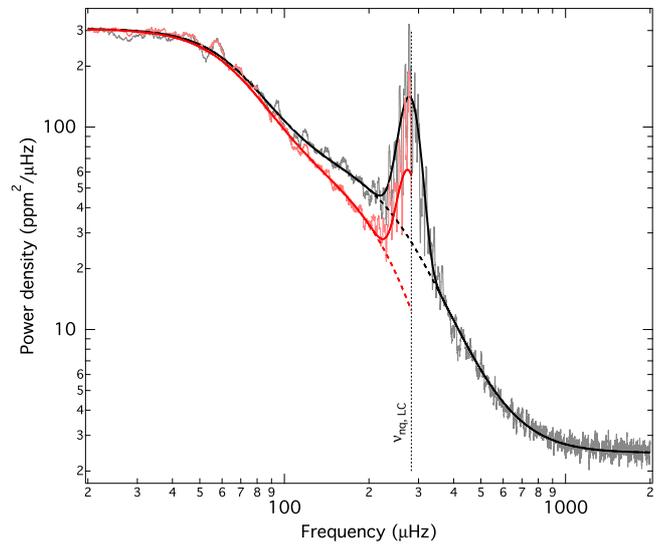}
	\caption{Heavily smoothed (5\mh\ boxcar filter) power density spectrum of the about 540 and 1000 day-long SC (grey line) and LC {\textit Kepler} time series (light-red line), respectively, of KIC\,5596656. The black lines correspond to the best-fit model for the SC data with (full line) and without (dashed line) the Gaussian component. The red lines indicate the SC model corrected for the amplitude suppression due to the longer sampling. The vertical dotted line marks the Nyquist frequency for the LC sampling.} 
	\label{fig:SCLC} 
	\end{center} 
\end{figure}

The variations we observe in stars are generally continuous but as soon as we measure them we naturally have to discretise them, i.e. the intrinsic signal is integrated for a certain time during each measurement. This is not a problem if the timescales of the variations are long compared to the integration time. For shorter timescales, however, the discretisation yields a partial cancelation of the signal, so that the resulting time series contains a damped signal. The damping is frequency dependent and becomes larger for increasing signal frequencies approaching the Nyquist frequency (given that the sampling is similar as the integration time, i.e. with little or no dead time). For the amplitude of a harmonic oscillation with a frequency $\nu$ this effect can be expressed by a damping factor,
\begin{equation}\label{eq:eta}
\eta = \mathrm{sinc} \left (\frac{\pi}{2}\frac{ \nu}{ \nu_\mathrm{nq}} \right ).
\end{equation}
Event though this is well known \citep[e.g.,][]{cha11}, the effect is often ignored, which can lead to a quite significant underestimation of the intrinsic amplitudes, especially for the \textit{Kepler} LC data. So it happens that, e.g., the amplitude of a mode at 200\mh\ is diminished by roughly 20\% ($\sim$35\% in power) in the LC data, which should no longer be ignored. 

The situation is similar for the granulation signal. What we observe for granulation is a quasi-stochastic signal covering a wide range of timescales with decreasing amplitudes for increasing frequencies. To measure the granulation amplitudes we are not so much interested in the amplitude at a particular frequency but in the integrated power at all timescales covered by the observations. Consequently, there is no simple correction factor for the signal damping as it is the case for oscillation modes and it was even unclear if the granulation signal is affected by the cancelation effects at all. In Fig.\,\ref{fig:SCLC} we show that it indeed is. It compares heavily smoothed power density spectra of one of the few cases where a red giant has been observed in LC as well as SC mode. Clearly, the power density of the LC time series (light-red line) drops faster towards higher frequencies than in the SC spectrum (grey line). To test that this is indeed due to the signal damping we computed a global model fit for the SC data (solid black line; see Sec.\,\ref{BGmodel}) and multiplied it with $\eta^2$ (red line). The resulting ``corrected'' model almost perfectly reproduces the LC spectrum. In reverse, we can therefore use Eq.\,\ref{eq:eta} to distort the global model during the fitting process in the same way as the intrinsic signal is distorted by the sampling in order to access the unperturbed granulation parameters. Ignoring the signal damping does barely affect most of the stars in our LC sample as their oscillation and granulation timescales are mostly well below the LC Nyquist frequency. But it introduces a systematically increasing underestimation of \textit{all} granulation parameters (also the timescales are affected) for stars with higher \num . 

Finally, we note that the white noise component of the power density spectrum is not affected by this sampling effect just like the binning of a light curve does not change the white noise level in the Fourier domain.

\section{Power density spectra modelling}	\label{BGmodel}

Modelling the power density spectrum of a solar-type oscillator appears to be straight forward but actually is a quite complex task. Before the \textit{Kepler} era, the background signal was mostly treated as a parasitic component one needs to get rid of quickly to access the oscillations. But the background signal itself includes interesting information on, e.g., the granulation or activity signal of the star. To extract this information most reliably, one needs an appropriate model whose complexity should be driven by the data quality. However, to our knowledge there is yet no consensus in the community about such a model and various approaches have been used in the past.  

Some early work was done by \citet{mat11}, who qualitatively compared granulation parameters that result from various background models for a large sample of red giants, observed by \textit{Kepler} for about 13 months. They found that even though the granulation parameters generally follow the same trends (e.g., as a function of \num ) there are significant differences between the different models. However, at the time the aim of the study was not to pick a ``best model'' nor were the data long enough to make firm conclusions. Now the time series are almost three times as long, enabling us to carry out a detailed comparison of different background models.

\subsection{The background model}

Originally, \citet{har85} used a function like $P(\nu) \propto 1/[1+ (\pi \nu \tau)^c]$ to model the solar background signal, where an exponent $c$ of 2 (i.e., a Lorentzian function) was adopted. The basic idea of this is that the granules cause a signal on the stellar surface that can be approximated by a sudden pulse with an exponential decay with a typical timescale $\tau$. The Fourier transform of such a pulse is a Lorentzian function. Later on it was shown (using better data) that an exponent of 4 is more appropriate (e.g. \citealt{mic09} for the Sun and \citealt{kal10b} for red giants) but also linear or exponential models have been used to describe the background underneath the oscillation signal. Note that a model with an exponent of 4 represents a simplification of $1/[1+ (\pi \nu \tau)^2]^{2}$, which is the Fourier transform of a symmetrically rising and decaying pulse. 

Another question is if more than one background component is required to sufficiently fit the observations. Each component is believed to represent a separate class of physical process such as stellar activity, different scales of granulation, or faculae, which are all strongly connected to the turbulent motions in the convective envelope. With the long time series observations of the Sun \citep[e.g., from SOHO/VIRGO;][]{fro97} it became evident that there is a slight depression in power density at about \num  /2 \citep[see, e.g.,][]{kal10b}, which can not be explained by a single background component in the vicinity of the oscillations. \citet{karoff12} interpreted this as the superposed signal of granulation and faculae and similar was found in other solar-type oscillators \citep{karoff13}. 

The reason why the exponents of Harvey-like models and the number of components needed to sufficiently reproduce the observed background are poorly determined is because one needs to access the high-frequency part of the spectrum (beyond the pulsation power excess, where the power rapidly drops). Until recently, this was only possible for the Sun and a few red giants with very good SNR but for most of the previously available data of solar-type oscillators, this part of the spectrum is hidden in the white noise. Apart from the Sun, however, the particular choice of the background model was not critical for pre-\textit{Kepler} data as the blurry background signal could be almost equally well modelled by different models. But with the increasing length of the \textit{Kepler} time series it becomes increasingly important to use an appropriate model that accounts for the complex structure of the background and that allows to access the oscillation modes in the best possible way. We have to keep in mind that the choice of the background model potentially influences the global parameters of the pulsation power excess (the effect is small but tilting the underlying background “redistributes” the excess power and shifts \num ) as well as the mode parameters itself (see Sec.\,\ref{sec:syseffects}). We therefore investigated different background models in more detail. 

In all cases, the power density spectra are modelled by the superposition of instrumental noise\footnote{which is either ``white'' or ``coloured''}, the contribution of one to three super-Lorentzian\footnote{Note that this function is often called Harvey or Harvey-like model. However, as the original Harvey model actually is a Lorentzian function we prefer the term super-Lorentzian} functions, and a power excess hump approximated by a Gaussian,
\begin{equation}\label{eq:bg}
P(\nu) = P_n' + \eta(\nu)^2 \left [ \sum_i \frac{\xi_i a_i^2 / b_i}{1+(\nu/b_i)^{c_i}} + P_g \exp{\frac{(\nu-\nu_\mathrm{max})^2}{2\sigma^2}} \right],
\end{equation}
where $P_n'$ corresponds to the instrumental noise contribution and $a_i$ and $c_i$ are the rms amplitude and exponent of the $i$-th background component, respectively. The parameter $b_i$ corresponds to the frequency at which the power of the component is equal to half its value at zero frequency and is called the characteristic frequency (which is equal to $(2\pi\tau)^{-1}$, with $\tau$ being the characteristic timescale). $P_g$, \num\ , and $\sigma$ are the height, the central frequency, and the width of the power excess hump, respectively. $\eta(\nu)$ is computed according to Eq.\,\ref{eq:eta} and distorts the background model in the same way as the observations are distorted by the sampling. The
factor $\xi$ is used to normalise $\int_0^\infty (\xi/b)/[1+(\nu/b)^c] d\nu = 1$ so that $a^2$ corresponds to the area under the super-Lorentzian function in the power density spectrum (which in turn is
equal to the variance in the time series that originates, for example, from granulation). Obviously, $\xi$ depends on $c$ and is 
for $c=2 \mapsto \xi = 2/\pi$,
for $c=4 \mapsto \xi = 2\sqrt{2}/\pi$. 
For $c$ different from an even integer (i.e., $c$ is a free parameter in the fit), the integral can't be solved analytically so that $\xi$ is undefined. We then fitted $A_i$ (i.e., the power density at zero frequency) instead of $\xi a_i^2/b_i$, and numerically integrated the fit to determine $a_i$.

In the current analysis we chose to use a frequency-dependent instrumental noise,
\begin{equation}\label{eq:bg1}
P_n' = P_n+  \frac{2\pi \alpha^2 / \beta}{1+(\nu/\beta)^2},
\end{equation}
where $P_n$ corresponds to the (constant) white noise and $\alpha$ and $\beta$ to the amplitude and characteristic frequency of the noise component introduced by the aperture optimisation. We note that neither a constant nor a coloured noise is affected by the sampling effect described in Sec.\,\ref{AmpDamp}, which is why $P_n'$ is not corrected for $\eta(\nu)$ in Eq.\,\ref{eq:bg}.

\subsection{Model comparison} \label{Sec:ModelComp}

To fit the models to the power density spectra we have used a Bayesian inference tool  \citep[\textit{MultiNest};][]{feroz09} assuming that the difference between the model and observed spectrum follow a $\chi^2$statistic with two degrees of freedom \citep{gabriel1994}\footnote{To ensure that the gap-filling has no (or only marginal) influence on this basic assumption we tested the residual signal in raw and gap-filled spectra for a number of stars and found no significant deviations.}. \textit{MultiNest} provides the posterior distributions for the parameter estimation as well as a realistic global model evidence. The big advantage of this global evidence compared to other statistical tools is that it is properly normalised and evaluated over the entire parameter space. It therefore allows to reliably rate how good a given model represents a given data set compared to another model with little risk to over-fit the data, which is often not given by a direct comparison of, e.g., the maximum likelihoods. This is because, roughly speaking, a more complex model (i.e., with more free parameters) tends to fit some data better (i.e., has a better likelihood) than a less complex model but in the Bayesian concept a model gets assigned a penalty for its complexity and needs to fit the data considerably better to get a higher model evidence than a less complex model. With this statistical measurement in hand, which tells us how good a given model represents the observations we compared the following models:

\begin{description}
\item[A:]the classical Harvey model with a single component and a fixed $c=2$,
\item[B:]a single super-Lorentzian function with $c$ fixed to 4,
\item[C:]a single component of the form $1/[1+(\nu/b)^2]^2$,
\item[D:]a single component with $c$ being a free parameter,
\item[E--H:] same as A--D but with two individual components
\end{description}

\begin{figure}[t]
	\begin{center}
	\includegraphics[width=0.5\textwidth]{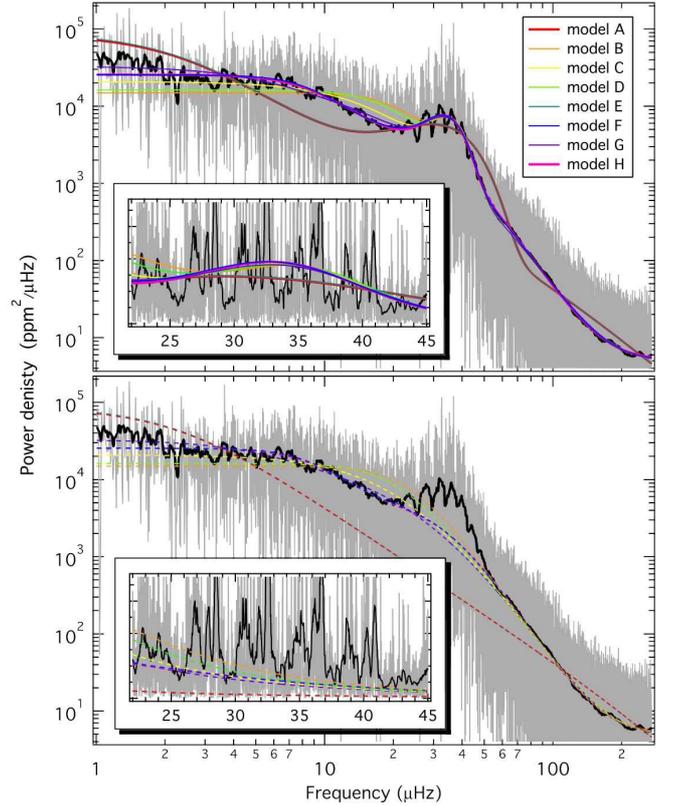}
	\caption{Raw (grey) and heavily smoothed (black) power density spectrum of KIC 7949599. Over-plotted in colour are the best fits model with (top panel) and without (bottom panel) the Gaussian component. Note that model A and E and model F and G effectively overlap. The inserts show enlarged sections around \num .} 
	\label{fig:bgmodelcomp} 
	\end{center} 
\end{figure}
\begin{figure}[t]
	\begin{center}
	\includegraphics[width=0.5\textwidth]{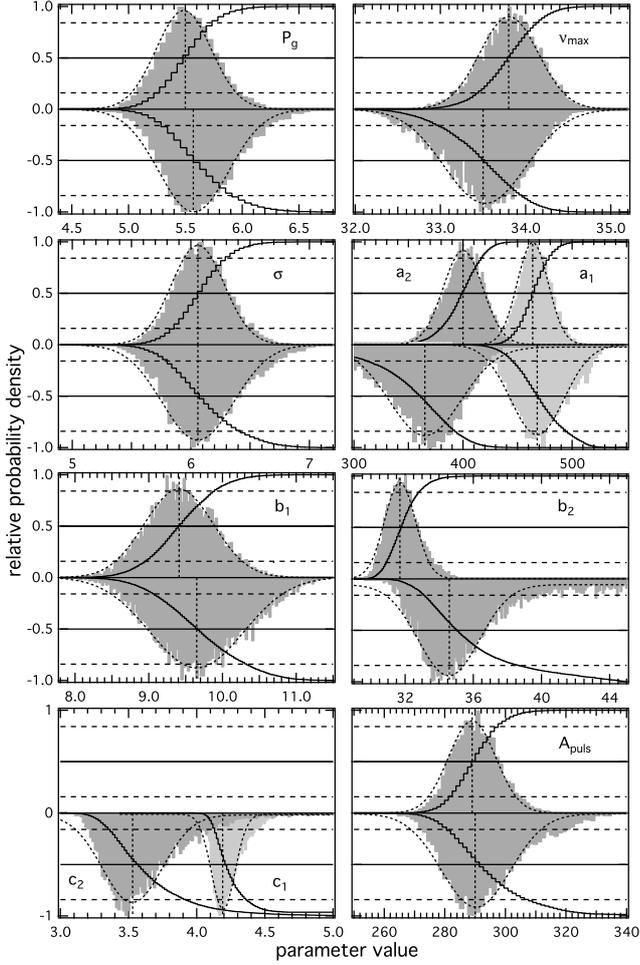}
	\caption{Histograms (grey bars) and cumulative distributions (black lines) of the probability density for the parameters determined for model F (pointing upwards) and model H (pointing downwards) for KIC 7949599, showing that all common parameters are practically equivalent (within the uncertainties). Horizontal solid and dashed lines indicate the median value and  the $\pm 1 \sigma$ limits (of a normal distribution), respectively. Vertical dotted lines mark the centre of a Gaussian fit to the histograms. The parameters units are the same as in Table\,\ref{tab:BGmodels}.} 
	\label{fig:probdens} 
	\end{center} 
\end{figure}

\begin{figure}[t]
	\begin{center}
	\includegraphics[width=0.5\textwidth]{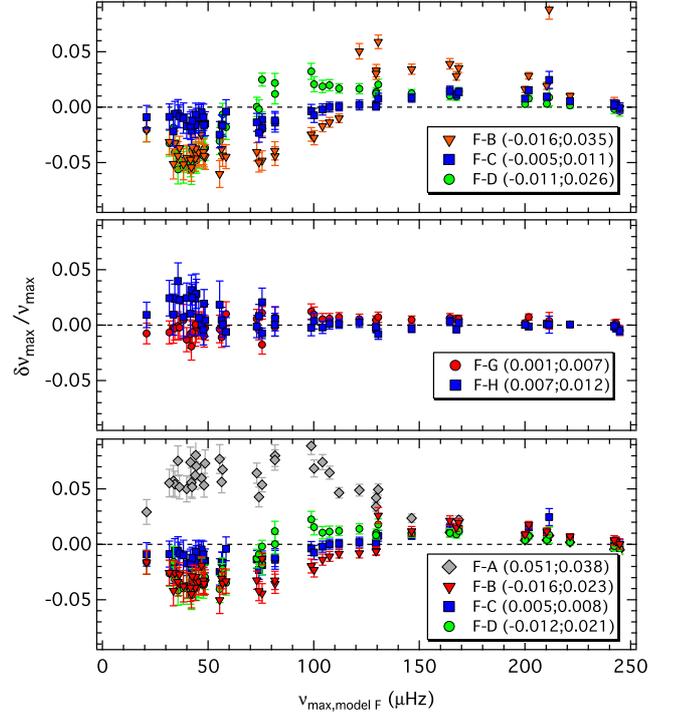}
	\caption{Relative differences in \num\ between model F and various other models for a sample of 50 red giants. The average relative difference and standard deviation is given in brackets. The top and middle panel show $\delta\nu_\mathrm{max}$ that result from fitting the entire spectrum (i.e., global fit) while in the bottom panel the fits are restricted to the range around the power excess hump (i.e., local fit). In both cases, there are clear systematic differences between 1- and 2-component models.} 
	\label{fig:numaxdiff} 
	\end{center} 
\end{figure}
To compare the different models we obviously have to assume that at least one model is true (i.e., the total probability is 1) and can then determine the individual model probabilities to $p_i = z_i/\sum_j z_j$, with $z$ being the global model evidence as delivered by \textit{MultiNest}. In Bayesian statistics, the model complexity is automatically covered by the multi-dimensional parameter volume that needs to be integrated over when computing $z$. This, however, depends also on the priors, as it is often the priors that determine the effective volume. Even tough \textit{MultiNest} uses per default uniform priors (i.e., no a priori knowledge is encoded) for all parameters, the particular choice of the parameter range that is evaluated influences $z$. The model evidence is not very sensitive to this (given that the posterior distributions are well covered), but to allow a ``fair'' model comparison, a given parameter, for example \num , needs to be evaluated over the same (or at least similar) range for all models.

\begin{table*}[t]
\begin{center}
\caption{Summary of the background model comparison for KIC 7949599. The model probabilities are determined according to $p_i=z_i / \sum_j z_j$, where $z$ corresponds to the global model evidence and $z_0$ is an abritrary reference value (ln($z_0$) = -138898). $P_g$ is given in 1000\ph\ $a_i$ in ppm, respectively. All frequency parameters are in \mh . An asterisks indicates that the parameter is kept fixed during the fitting. Last digit uncertainties are given in parenthesis. 
\label{tab:BGmodels}}
\begin{tabular}{c|rr|ccc|ccc|ccc}
\hline
\hline
\noalign{\smallskip}
&&&\multicolumn{3}{c|}{Gaussian}&\multicolumn{3}{c|}{1$^{st}$ component}&\multicolumn{3}{c}{2$^{nd}$ component}\\
 & ln($z/z_0$) & $p$ & $P_g$ & \num & $\sigma$ & $a_1$ & $b_1$ & $c_1$ & $a_2$ & $b_2$ & $c_2$ \\
\noalign{\smallskip}
\hline
\noalign{\smallskip}
A  &  -1587.7 & $<10^{-200}$  & 5.4(2)   &    30.38(02)  &     13.1(2)  &     560(12)   &    2.3(1) &  									2$^*$    		&&&		\\
B  &  -255.7&   $\sim 10^{-111}$   & 4.8(3)   &    35.7(3)  &      5.1(2)   &    624(6)    &   23.7(2) &  									4$^*$    		&&&		\\
C  &   -75.8 & $\sim 10^{-33\,\,\,}$             & 5.5(3)   &    34.5(2)  &     6.0(1)   &    606(6)    &   22.5(2) &  									2/4$^*$    		&&&		\\
D  &   -243.4 & $\sim 10^{-102}$             & 5.1(3)   &    35.2(2)  &     5.7(2)   &    601(28)  &	20.8(4)  & 									3.7(1)            &&& 		\\
\noalign{\smallskip}
\noalign{\smallskip}
E  &  -1592.4 & $<10^{-200}$  & 5.4(2)   &    30.42(02)  &     13.2(2)   &    571(15)    &   2.3(2)  &  2$^*$&  31(4)  &   34.1(6)	  		&2$^*$	   		\\
\bf{F}  &  \bf{-1.7} & \bf{0.166}              & \bf{5.5(2)}   &    \bf{33.8(4)}  &     \bf{6.1(2)}   &    \bf{466(14)}   &   \bf{9.4(5)}  &  \bf{4}$^*$&  \bf{399(19)}     &		\bf{31.9(1)}  			&\bf{4}$^*$		\\
G  &  -36.6 & $\sim 10^{-16}$             & 5.7(2)   &    33.9(2)  &     6.4(2)   &    352(26)   &   8.5(9)  &  2/4$^*$&  502(18)      &	25.7(6)		&	2/4$^*$  	\\
\bf{H}  &  \bf{-0.1} & \bf{0.833}             & \bf{5.6(3)}   &    \bf{33.5(5)}  &     \bf{6.1(3)}   &    \bf{470(35)}  &	\bf{9.7(6)}   & \bf{3.6(3)}&  \bf{365(59)}       	  &    \bf{35.8(3)}	&\bf{4.2(2)}\\
\noalign{\smallskip}
\hline
\end{tabular}
\end{center}
\end{table*}

We tested these models for 10 different red giants covering various \num\ and exemplarily list the model probabilities and best-fit parameters for KIC 7949599 in Tab.\,\ref{tab:BGmodels}, for which the model comparison gives a clear picture. This is also typical for the other stars. The first interesting result is that the original Harvey models (A and E) are completely ruled out as they can not reproduce the observed spectrum at all. This is not surprising as can be seen in Fig.\,\ref{fig:bgmodelcomp}, where we show the observed spectrum along with some of the model fits. Obviously, the slope in the high-frequency part of the spectrum, above the power excess, is much too steep to be reproduced by a model with an exponent of 2. Also interesting is that none of the models with a single background component (A to D) gets assigned a model probability that comes even close to those of the 2 component models. One can therefore safely assume that the background signal in red giants consists of multiple components, which is in agreement with what was found for main-sequence stars \citep[e.g.,][]{karoff13}. Among the eight tested models there are only two that result in a high enough probability to be considered a good representation (in our sample of models) of the data. Even though there is a preference for model H the probability contrast of about 5 between model F and H is not enough to provide significant evidence for one particular choice. According to \citet{jeffreys61} an odds ratio of up to 3:1 is called weak evidence, only for $>$10:1 one can speak of strong evidence. In fact, the best-fit exponents of model H are close to 4 (within $\sim$1.3$\sigma$) but more important here is that model F and H result in similar parameters (actually identical within the uncertainties). This can be seen from Fig.\,\ref{fig:probdens}, where we plot histograms of the posterior probability density for all parameters of model F (upwards) and H (downwards) along with their cumulative distributions. Obviously, the posterior distributions for all (common) parameters strongly overlap with a tendency that model H results in broader distributions, i.e. larger uncertainties. The models can therefore be considered practically equivalent. Fig.\,\ref{fig:probdens} also illustrates that the posterior distributions are not always Gaussian. It can therefore be misleading to define the best-fit parameter and its uncertainty as the average value and its standard deviation (which is equivalent to fitting a Gaussian to the distribution). A more general definition for the best-fit parameter and its uncertainty limits is the value at which the cumulative distribution is equal to 0.5 (i.e., the median) and 0.5$\pm$0.3414 (which is equivalent to $\pm$1$\sigma$ uncertainties in case of a Gaussian ), respectively. Note that, even though the uncertainties can be asymmetric we provide only a single average value for the sake of simplicity. 

The result of our model comparison would have been different if we would have used the classical likelihood ratio test. The likelihood computed from the difference between the best-fit models and the observed spectrum, clearly favours model H with an odds ratio of about 99:1, which is, as shown by the Bayesian analysis clearly over-interpreting the data. Or with other words, the data do not (yet) support to explicitly determine the exponent of the background model. 

The main conclusion for the other 9 test stars is similar and we do not find a case that differs significantly from that of KIC 7949599. For some stars in this representative sample model H is the most probable, for some model F is but for none of them the probability contrast is large enough to strongly support a specific choice. We do also not find any indication for a correlation between the specific model choice and the stellar properties like the surface gravity. Even though model H might be more realistic in some cases, we found no compelling argument to prefer model H over model F and since model F gives practically the same background parameters, is more robust (as it has less free parameters), and tends to result in smaller uncertainties, we decided to use model F for the subsequent analysis.

\begin{figure}[t]
	\begin{center}
	\includegraphics[width=0.5\textwidth]{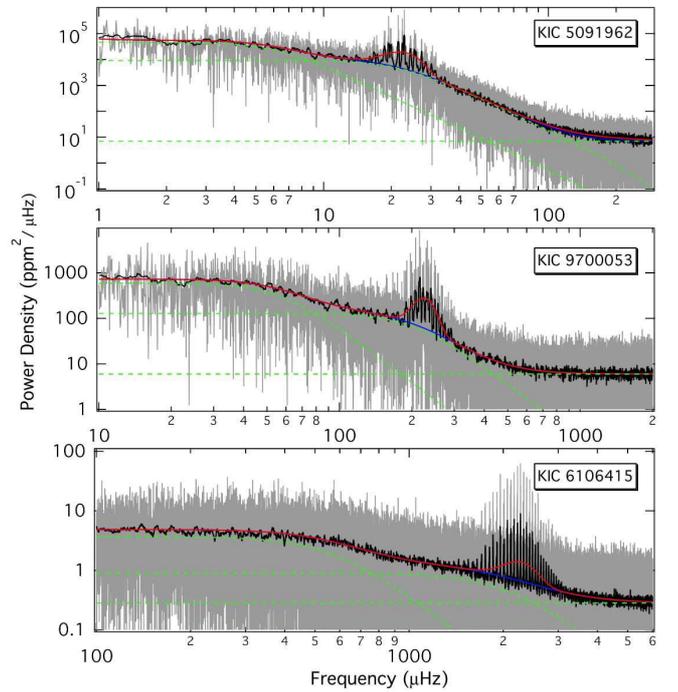}
	\caption{Power density spectra of three typical stars with $\nu_\mathrm{max} \simeq$ 22, 220, and 2200\mh , respectively, showing that all timescales and amplitudes (granulation as well as pulsation) scale simultaneously. Grey and black lines indicate the raw and heavily smoothed spectrum, respectively. The global fit is shown with (red) and without (blue) the Gaussian component. Green lines indicate the individual background and white noise components of the fit.} 
	\label{fig:FouSample} 
	\end{center} 
\end{figure}

\subsection{Systematic effects on \num} \label{sec:syseffects}

The global model fit does not only allow to determine the background and global oscillation parameters but also provides an estimate for the background signal underneath the pulsation modes, which is an important ingredient for the subsequent analysis of the oscillations. As can be seen from the insert in Fig.\,\ref{fig:bgmodelcomp} it is difficult to decide which model gives the most realistic background as we can not easily disentangle the pulsation from the granulation signal and all models (apart from the classical Harvey model) seem to estimate the background equally well. However, there are small differences, especially at the low-frequency end of the power excess. These differences redistribute the power in the hump and potentially lead to different estimates for \num . We therefore selected 50 red giants with a low white noise component (where about half of the stars with \num\ below 70\mh\ were chosen to be RC stars) and compare \num\ that results from model F and various other models in Fig.\,\ref{fig:numaxdiff}. Whereas 2-component models give roughly the same \num\ (see middle panel) with only minor systematic deviations, 1-component models result in \num\ estimates that clearly differ from those determined with model F (see top panel). Obviously there is a systematic trend as a function of \num , so that \num\ from 1-component models are systematically over- and underestimated for stars with low and high \num , respectively. The quite significant deviations of up to $\sim$6\% disagree to some extend with what was found by, e.g., \citet{hek12} who argued that \num\ from different fitters (using different background models) is generally in good agreement. An explanation for that might be that the individual deviations cancel out when looking at the entire sample as a whole. The mean difference of the sample in Fig.\,\ref{fig:numaxdiff} is always within the typical uncertainties of the individual measurements. Another reason could be that the methods mentioned in \citet{hek12} that use only one background component do not try to fit the whole spectrum, but only the region around the pulsation power excess. They are therefore less sensitive to the overall structure of the spectrum (and not suitable to deliver granulation parameters) and might give a better estimate of \num\ than in our analysis, where we needed to fit the whole spectrum to allow the model comparison. However, to test if the situation changes in a local analysis we redid the fits for the range 0.5 to 5 times \num\ and show the results in the bottom panel of Fig.\,\ref{fig:numaxdiff}. Note that for model A we had to trim the spectrum even more to [0.5 - 2] $\times$ \num , as it is not possible to follow the rapid decay of the granulation signal at high frequencies with the classical Harvey model. Obviously also a local treatment of the background can systematically effect the determination of \num\ although the effect is reduced. We therefore advise against 1) the use of only one background component (even in a local analysis), and 2) the use of the classical Harvey model. A 1-component Harvey-like model might be applicable if the instrumental white noise is high enough so that the high-frequency part of the granulation signal is hidden therein. For stars like the one shown in the bottom panel of Fig.\,\ref{fig:FouSample}, where the contrast between the low-frequency granulation signal and the white noise is only about 10:1 (compared to about $10^4$:1 for the star in the top panel) the functional form and number of background components become less important. In an extreme case with the white noise being at a similar level as the granulation signal even a straight line (i.e, a white component) will do.

The above analysis shows that a different treatment of the background systematically effects the determination of \num , but it does not reveal which of the background models gives the correct \num . This question can only be answered with simulated data set with the actual \num\ known, which we try in the following. As an input for the pulsation signal we use the $l$ = 0, 1, and 2 adiabatic frequencies\footnote{computed with Guenther's nonradial nonadiabatic stellar pulsation program; \citet{gue94}} of a 1.25\,M\sun\ near solar-calibrated YREC \citep[Yale Stellar Evolution Code;][]{gue92,dem08} model with a radius and effective temperature chosen so that \num\ is close to 50\mh\ (where we find the largest systematic deviations, see Fig.\,\ref{fig:numaxdiff}). For details about the constitutive physics we refer to \citet{kal10c} and references therein. Mode amplitude ratios are modulated between 0 and 1 by a Gaussian centred on 50\mh\ (and a width of 8\mh ), where we assume a mode visibility of 1.0, 1.5, and 0.5 for $l$ = 0, 1, and 2 modes, respectively. The amplitude ratios of $l$ = 1 and 2 modes are additionally modulated according to their inverse mode inertia \citep[i.e., modes with a low inertia have large amplitudes and vice versa; see, e.g.,][]{dup09}. The $l$ = 0 and 2 modes are assumed to have a constant mode lifetime of 30 and 40 days, respectively, and the lifetime of $l$ = 1 modes is modulated between 100 and 200 days according to their mode inertia. To compute the artificial 1142 day-long time series (sampled each $\sim$5.9 minutes, see below) for this set of solar-like modes we add up damped and randomly re-excited oscillations, generated by the method described by \citet{cha97}. In order to get realistic absolute amplitudes we rescale the artificial time series so that its rms scatter is equal to 200\,ppm (which corresponds to the total pulsation amplitude, see Sec.\,\ref{sec:apuls} below) and add random scatter so that the white noise in the resulting power density spectrum is equal to 3\,ppm$^2/\mu$Hz. Since the modes in this simulated power excess are not symmetrically distributed within the power excess, the real \num\ does not have to correspond to input value of the centre. To check this, we fit a Gaussian to the simulated power excess and find the power to be symmetrically distributed around $\nu_\mathrm{max,ref}$ = 49.61$\pm$0.06\mh . We note that in reality the shape of the pulsation power excess is not necessarily Gaussian but as long as the power is distributed roughly symmetric \citep[and there is no indication that it is not, see, e.g.,][]{kal10a} any symmetric function will give a good estimate of the intrinsic centre of the power excess. 

For the granulation background we add two sets of random symmetric exponential pulses, with decay times that correspond to characteristic frequencies, $b_i$ = 14 and 46\mh . The characteristic amplitudes $a_i$ are set to 250\,ppm by rescaling the time series to the corresponding rms scatter. To also account for the sampling effect described in Sec.\,\ref{AmpDamp}, the original time series is oversampled by a factor of 5 and finally re-sampled to the LC sampling by averaging.

We then analyse the simulated spectrum between 10\mh\ and the Nyqvist frequency. For the 2-component models, model F results in a \num\ = 0.995$\pm$0.01$\nu_\mathrm{max,ref}$ and is therefore in excellent agreement with the reference value. Furthermore all other input parameters are reproduced to within $\sim$2$\sigma$. Similar can be said for model G and H. Only for model E, the peak frequency is grossly underestimated to \num\ = 0.94$\pm$0.01$\nu_\mathrm{max,ref}$, even though we truncate the spectrum to [0.5 - 2] $\times$ \num\ to get a useful fit. As we expected, the 1-component models yield peak frequencies that are significantly different from the reference value. While model A underestimates \num\ even more (0.91$\nu_\mathrm{max,ref}$), model B, C, and D overestimate the peak frequency to \num\ = 1.03, 1.02, and 1.02$\nu_\mathrm{max,ref}$, respectively, where the uncertainties are always on the order of 1\%.

As mentioned earlier, these deviations are presumably due to the fact that an inappropriate background model redistributes the power in the power excess and shifts the centre. If so, the effect should become smaller if the shape of background signal is less distinctive, i.e. large parts of it are hidden in the white noise. This is indeed what we observe. If we increase the white noise in our simulated spectrum by a factor of 200 (which dilutes the contrast between the low-frequency granulation signal and the white noise to about 10:1) the deviations between the various models become less pronounced. While we get almost exactly the same \num\ as before for model F, all other models result in a \num\ that is within [0.97-1.0]$\times \nu_\mathrm{max,ref}$. Interestingly, increasing the white noise by such a large amount has only small effects on the uncertainties of \num , which are now about 50\% larger than before.  

This test with a simulated spectrum seems to supports our conclusions from above, namely that a model with two super-Lorentzian functions is suited best to reproduce the observed granulation signal and to determine \num . Neither a 1-component model nor a model with an exponent significantly different from four should be used (or at least only under certain circumstances; see above). For completeness we also simulated a star with a much higher \num\ = 150\mh\ and find the peak frequencies from the various models to more or less converge to the same value, which is in agreement with what can be seen in Fig.\,\ref{fig:numaxdiff}.

\subsection{Fitting the sample}
As mentioned above we fit model F to our sample of LC and SC data.  In addition to the two components in the vicinity of the power excess, we apply a third component to account for long-period signal intrinsic to the star (e.g., due to activity) with periods that are usually much longer than 10/\num , but also residual instrumental signal. From previous work \citep[e.g.,][]{mat11} it is know that basically all (intrinsic) parameters scale with \num , for which we have already good initial guesses from, e.g., \citet{kal12}. To optimise the computational performance we tried to keep the parameter ranges relatively narrow. During the fit \num\ is kept within $\pm$10\% of its initial guess. The characteristic frequencies $b_i$ are allowed to vary between 0.01 to 0.3, 0.1 to 0.7, and 0.7 to 2 times the initial guess of \num . All amplitude parameters ($a_i$ and $\alpha$) are kept below the rms scatter of the time series (i.e., the square root of the integrated power in the spectrum). The height and width of the Gaussian are allowed to vary between 0 to half the maximum power in the spectrum and 0.05 to 0.3 times the initial guess of \num , respectively, and $\beta$ may vary between 0 and the Nyquist frequency. \textit{MultiNest} then delivers the posterior probability distributions for the 12 free parameters. We check if they are well sampled within the parameter range and finally marginalise them to determine the most probable parameters and their 1$\sigma$ uncertainties. One parameter that is not directly fitted but evaluated in the subsequent analysis is the so-called total pulsation amplitude $A_\mathrm{puls}$. It is supposed to represent the entire pulsation energy in the power excess, i.e., the sum of \textit{all} individual mode amplitudes. During the fitting procedure we compute $A_\mathrm{puls}$ as $(\sigma P_g \sqrt{2\pi})^{1/2}$, which is the square root of the area under the Gaussian component of our model, for each evaluated combination of $P_g$ and $\sigma$ to build a posterior distribution from which we determine the most probable value of $A_\mathrm{puls}$ and its uncertainty.

The long \textit{Kepler} time series allowed us to determine \num\ on average to within about 0.9\%, the characteristic granulation frequencies $b_1$ and $b_2$ to within about 5.3 and 3.6\%, respectively, the granulation amplitudes $a_1$ and $a_2$ to better than 3.2 and 5.2\%, respectively, and $A_\mathrm{puls}$ to within about 2.8\% on average.

In Fig.\,\ref{fig:FouSample} we show the power density spectra of three representative stars, whose \num\ differ by a factor of $\sim$10 from one to the next, along with the best-fit model (F) and its individual components.

\begin{table*}[t]
\begin{center}
\caption{Parameters of power law fits to the granulation. The surface gravity $g$ is in \textit{cgs} units, frequency parameters (\num , $b_1$, and $b_2$) are in \mh , $\tau_\mathrm{eff}$ is in seconds, and amplitude parameters ($a$, $A_\mathrm{gran}$, and $A_\mathrm{puls}$) are in ppm. The stellar mass $M$ is in solar units and $T = T_\mathrm{eff}/5777$. $\sigma$ indicates the rms variations of the residuals in percent. If a coefficient is left blank means that the corresponding parameter is not included in the fit. Last digit uncertainties are given in parenthesis. $\rho_{Y_i,X_i}$ is the magnitude of the correlation coefficient (in percent) between $X_i=\log_{10}(x_i)$ and $Y_i=\log_{10}{y_i}$, with $y_i = y_1/\prod_{j=0}^{i-1}x_j = k\,x_i^c$, with $x_0=1$. The index $i$ indicates the independent variable (from left to right) of the corresponding power law fit.
\label{tab:fitpar}}
\begin{tabular}{l|c|c|cc|cc|c}
\hline
\hline
\noalign{\smallskip}
&$b_1=k \,\nu_\mathrm{max}^s$ & $b_2=k \,\nu_\mathrm{max}^s$ & \multicolumn{2}{c}{ $a=k \,\nu_\mathrm{max}^s M^t$}&\multicolumn{2}{|c|}{$\tau_\mathrm{eff}=k\,\nu_\mathrm{max}^s= k\,g^s\,T^t$}&$A_\mathrm{gran}=k\,\nu_\mathrm{max}^s$\\
\noalign{\smallskip}
\hline
\noalign{\smallskip}
$k$&0.317(2)&0.948(3)&3382(9)&3710(21)&836(4)&2.00(4)$\cdot 10^6$&3335(9)\\
$s$&0.970(2)&0.992(2)&-0.609(2)&-0.613(2)&-0.886(2)&-0.853(3)&-0.564(2)\\
$t$&              &             &               &-0.26(3)&&-0.41(5)&\\
$\sigma$&10.2&8.7&16.5&14.4&11.0&9.3&13.5\\
$\rho_{Y_i,X_i}$&99&99&95&44&99&99,52&98\\
\noalign{\smallskip}
\hline
\noalign{\smallskip}
&\multicolumn{3}{c|}{$A_\mathrm{gran}$=$k \, g^s M^t T^u$}&$A_\mathrm{puls}$=$k\, g^s M^t T^u$&\multicolumn{3}{c}{$A_\mathrm{gran} = k \, A_\mathrm{puls}^s T^t = k\,A_\mathrm{puls}^s g^t M^u$}\\
\noalign{\smallskip}
\hline
\noalign{\smallskip}
$k$&\multicolumn{1}{c}{9799(43)}&\multicolumn{1}{c}{10887(127)}&\multicolumn{1}{c|}{11054(385)}&9344(348)&\multicolumn{1}{c}{4.57(3)}&\multicolumn{1}{c}{4.79(4)}&36(3)\\
$s$&\multicolumn{1}{c}{-0.549(6)}&\multicolumn{1}{c}{-0.555(6)}&\multicolumn{1}{c|}{-0.556(4)}&-0.657(4)&\multicolumn{1}{c}{0.855(3)}&\multicolumn{1}{c}{0.837(3)}&0.63(1)\\
$t$&           \multicolumn{1}{c}{}              &\multicolumn{1}{c}{-0.24(2)}&\multicolumn{1}{c|}{-0.24(3)}&-0.35(3)&\multicolumn{1}{c}{}&\multicolumn{1}{c}{0.24(3)}&-0.15(1)\\
$u$&	\multicolumn{1}{c}{}			&\multicolumn{1}{c}{	}	&\multicolumn{1}{c|}{0.05(9)}&0.04(8)&\multicolumn{1}{c}{}&\multicolumn{1}{c}{}&-0.06(1)\\
$\sigma$&\multicolumn{1}{c}{13.1}&\multicolumn{1}{c}{11.0}&\multicolumn{1}{c|}{10.7}&13.6&\multicolumn{1}{c}{8.0}&\multicolumn{1}{c}{10.3}&7.3\\
$\rho_{Y_i,X_i}$&\multicolumn{1}{c}{97}&\multicolumn{1}{c}{44}&\multicolumn{1}{c|}{80}&97,41,76&\multicolumn{1}{c}{99}&\multicolumn{1}{c}{57}&99,91,25\\
\noalign{\smallskip}
\hline
\end{tabular}
\end{center}
\end{table*}

\section{Scaling relations}	\label{sec:scaling}

\subsection{Granulation timescales (or frequencies)} \label{Sec:GranFreq}
Using a few basic physical assumptions one can estimate the granulation timescales and amplitudes from the stellar parameters. For example, \citet{kjeldsen11} and \citet{mat11} argued that convection cells cover a vertical distance that is proportional to the atmospheric pressure scale height, $H_p\propto T_\mathrm{eff}/g$, at a speed approximately proportional to the speed of sound, $c_s \propto \sqrt{T_\mathrm{eff}}$, where $g$ is the surface gravity and $T_\mathrm{eff}$ is the effective temperature. Therefore the characteristic timescale of granulation or more conveniently the characteristic frequency (since we measure them in a frequency spectrum) can be expressed as $\nu_\mathrm{gran} \propto c_s/H_p \propto g/\sqrt{T_\mathrm{eff}} \propto \nu_\mathrm{max}$, or with other words there should be a tight relation between the characteristic granulation frequency and \num\ that should be linear in a first approximation. This is indeed what we observe. The top panel of Fig.\,\ref{fig:ab_numax} shows that there is a tight relation between the characteristic frequencies ($b_1$ and $b_2$) and \num , covering a range in \num\ of more than 3 orders of magnitude from the Sun to high up the giant branch. A power law fit (black lines) reveals that both relations are almost linear as both exponents are close to 1. The best fit coefficients and their uncertainties are listed in Tab.\,\ref{tab:fitpar}. Furthermore we find that the dispersion along this relation is with about 10 and 9\% only about twice as large as the average uncertainties of the individual measurements. This indicates that a power law is too simplistic but also that other properties (like the evolutionary state, chemical composition, magnetic activity, etc.) play only a minor role. 

The uncertainties of the fit coefficients might appear unrealistically small but we note that they are based on uncertainties in $b_i$ as well as \num . For the fit we again used \textit{MultiNest}, which can only consider uncertainties in the dependent variable (as most other fitting algorithms). To account also for the uncertainties of the independent variable we add normally distributed uncertainties to \num\ (with $\sigma$ being equal to the actual uncertainties of the individual measurements) and redo the fit adding up the probability density distributions. We iterate this procedure until each fitting parameter converges to a value that stays within one fifth of the corresponding uncertainty.  Additionally we note that the high correlation coefficients (see Tab.\,\ref{tab:fitpar}) between $b_i$ and \num\ indicate that one can accurately constrain the power-law parameters.

\subsection{Granulation amplitudes} \label{Sec:GranAmp}
\begin{figure}[t]
	\begin{center}
	\includegraphics[width=0.5\textwidth]{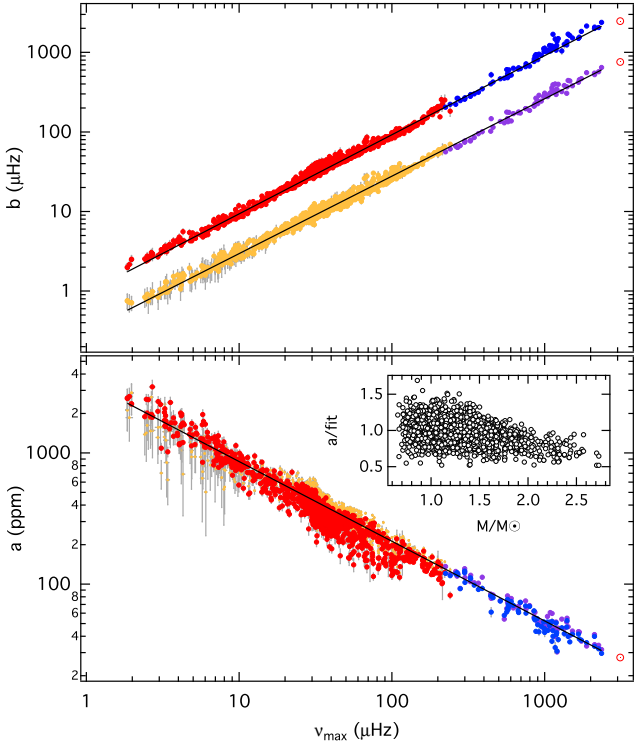}
	\caption{Characteristic frequencies (top) and rms amplitudes (bottom) as a function of \num\ for our sample of LC (red/orange) and SC (blue/magenta) stars. Orange and magenta coloured symbols correspond to the lower frequency component ($a_1$ and $b_1$). Black lines indicate power law fits (see Tab.\,\ref{tab:fitpar}) and red open circles solar values. The insert shows the rms amplitudes of the LC sample divided by the power law fit as a function of stellar mass.} 
	\label{fig:ab_numax} 
	\end{center} 
\end{figure}

Using similar assumptions as above, \citet{kjeldsen11} argued that the granulation power at \num\ scales as $P_\mathrm{gran}(\nu_\mathrm{max}) \propto \nu_\mathrm{max}^{-2}$. However, $P_\mathrm{gran}$ is not really qualified to serve as an ``amplitude'' parameter for granulation as it gives the power at a certain frequency and not, e.g., a proxy for the total energy that is contained in the granulation signal. More appropriate is the rms amplitude $a$, which is defined as the square root of the integrated granulation signal in the power density spectrum and represents the rms intensity fluctuations in the time series that are due to granulation. From Eq.\,\ref{eq:bg} follows that $P_\mathrm{gran}(\nu_\mathrm{max}) \propto (a^2/b)/[1+(\nu_\mathrm{max}/b)^c]$. With $P_\mathrm{gran}(\nu_\mathrm{max}) \propto \nu_\mathrm{max}^{-2}$ and $b \propto \nu_\mathrm{max}$ we then infer that $a \propto \nu_\mathrm{max}^{-0.5}$. Basically this comes from the fact that the intensity fluctuations scale as the inverse of the square root of the number of granules on the stellar surface \citep[e.g.][]{ludwig2006}. This is again similar to what we observe. The bottom panel of Fig.\,\ref{fig:ab_numax} illustrates the rms amplitude of both granulation components ($a_1$ and $a_2$) as a function of \num\ showing that both components have comparable amplitudes that are tightly correlated to \num . A power law fit (green line) to both amplitudes reveals that $a \propto \nu_\mathrm{max}^{-0.61}$, which is close to what we expected (based on some very basic assumptions) and to what has been observed by others \citep[e.g.,][who fitted the variance of a sample of \textit{Kepler} red giants, which is dominated by the granulation signal]{hek12}. In contrast to the granulation frequency scaling, the dispersion along the fit is with about 16\% significantly larger (three to five times) than the average uncertainties of the individual measurements. This indicates that the granulation amplitudes are primarily correlated to \num\ (which in turn is dominated by the stellar radius) but also other parameters like the stellar mass and/or the chemical composition potentially have a significant influence on the granulation amplitude \citep[see, e.g.,][]{mos12a}. To test this we correct the granulation amplitudes for the dependency on \num\ and show the residuals as a function of stellar mass (insert in the bottom panel of Fig.\,\ref{fig:ab_numax}). Stellar fundamental parameters have been computed with the method described by \citet{kal10b}, where we use \num , $\Delta\nu$, and $T_\mathrm{eff}$ as an input for the LC sample and \num\ and $T_\mathrm{eff}$ for the SC sample. Effective temperatures are taken from the \textit{Kepler} Input Catalog \citep{brown11} and corrected according to \cite{thy12}. 

Obviously there is a correlation with mass so that high-mass stars tend to have smaller amplitudes than low-mass stars. To account for this additional mass dependency (\num\ already depends on mass) we add a term and fit $a = k \nu_\mathrm{max}^s M^t$ instead of $a = k \nu_\mathrm{max}^s$, which results in a considerably better fit (the global evidence of the \textit{MulitNest} fit is many orders of magnitudes better). The best fit coefficients and their uncertainties (determined with the procedure described in Sec.\,\ref{Sec:GranFreq}) are listed in Tab.\,\ref{tab:fitpar}. 

\subsection{Characteristic timescale, $\tau_\mathrm{eff}$}
For the Sun, any background signal between a few 100 to a few 1000\mh\ is usually attributed to granulation \citep[see e.g.][for a summary of various interpretations found in the literature]{karoff13}. A major problem in this frequency range results from a kink that is visible in the background spectrum just below the oscillation power excess. This feature has been first identified in the solar irradiance data from the SOHO/VIRGO instrument \citep[e.g.,][and references therein]{mic09} but has meanwhile been observed by \textit{Kepler} in other stars ranging from red giants \citep[e.g.,][]{mat11} to main-sequence stars \citep[e.g.,][]{karoff13}. The depression in power at about $\nu_\mathrm{max}/2$ (see Fig.\,\ref{fig:FouSample}) arises from the fact that the background signal in the vicinity of the oscillation signal consists not of one but two statistically significant (see Sec.\,\ref{Sec:ModelComp}) individual components. One component certainly can be attributed to granulation but the physical origin of the second component is not yet clear and still subject of debates \citep[see e.g.,][for a summary]{sam13a, sam13} and is therefore missing in theoretical models. Thus a direct comparison between our measured granulation parameters and those from theoretical models is difficult as they measure different things.

\begin{figure}[t]
	\begin{center}
	\includegraphics[width=0.5\textwidth]{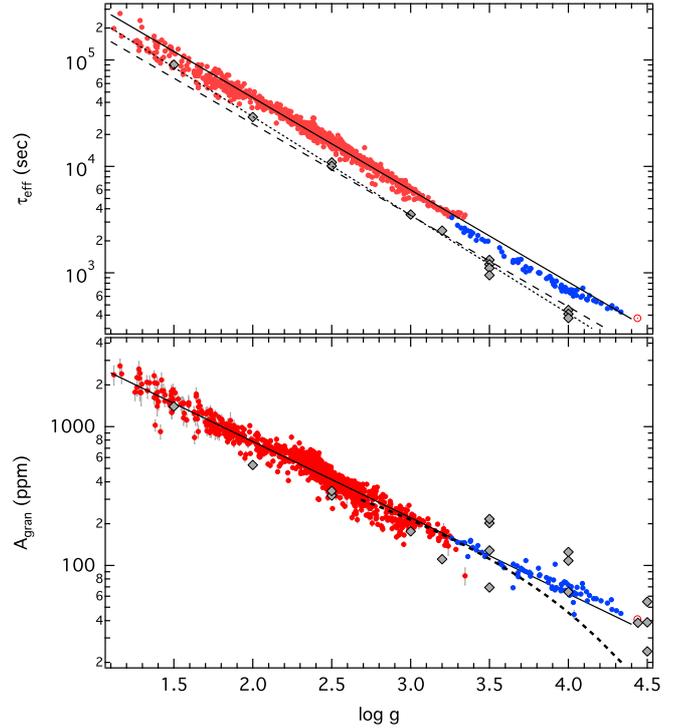}
	\caption{Characteristic timescale $\tau_\mathrm{eft}$ (top) and intensity fluctuations $A_\mathrm{gran}$ (bottom) as a function of \lg\ for our sample of LC (red) and SC (blue) stars. Solid lines correspond to power law fits. The dashed line in the top panel corresponds to a power law fit to effective timescales obtained from 3D simulations \citep{mat11}.  Diamond symbols indicate values obtained from the granulation spectrum of 3D hydrodynamical models \citep{sam13} and the dotted line corresponds to a power law fit to these models. The dashed line in the bottom panel shows the relation between the 8hr-flicker amplitude and \lg .} 
	\label{fig:efoldtime} 
	\end{center} 
\end{figure}

To still be able to compare characteristic granulation timescales obtained from models and observations (or also from observations using different descriptions of the granulation spectrum) one can define an effective timescale $\tau_\mathrm{eff}$ as the e-folding time of the autocorrelation function (ACF) of the time series \citep[e.g.,][]{mat11}. The characteristic timescale of any description of the granulation signal in a power spectrum measures the temporal correlation of the signal in the time domain, which can also be expressed by the width of the ACF. Recalling that the autocorrelation of a signal corresponds to the Fourier transform of its power spectrum, we can determine $\tau_\mathrm{eff}$ of our two-component granulation model by computing the time where the Fourier transform (i.e., the ACF) of the two superposed granulation components drops below $e^{-1}$. The result is shown in the top panel of Fig.\,\ref{fig:efoldtime}, where we choose to plot $\tau_\mathrm{eff}$ as a function of \lg\ instead of \num, as the surface gravity is a more basic stellar property and easy to determine for the models. 

From a theoretical point of view and with 3-D modelling, \citet{sam13} suggested that the characteristic granulation timescale can be described as $\tau_\mathrm{eff} \propto (\nu_\mathrm{max} \mathcal{M}_a)^{-1}$, where $\mathcal{M}_a$ is the turbulent Mach number in the photosphere. The Mach number is difficult (or even impossible) to determine observationally but the authors also found that it approximately scales as $\mathcal{M}_a \propto T_\mathrm{eff}^{2.4}/g^{0.15}$. With this and with $\nu_\mathrm{max} \propto g/\sqrt{T_\mathrm{eff}}$ the above theoretical description translates to $\tau_\mathrm{eff} \propto g^{-0.85}\,T_\mathrm{eff}^{-1.9}$. A power-law fit to the observation reveals,
 \begin{equation} \label{eq:taueff}
\tau_\mathrm{eff} \propto g^{-0.85}\,T^{-0.4},
\end{equation}
where $T = T_\mathrm{eff}/5777\,K$. Interestingly, the scaling with $g$ is exactly as expected. Only the temperature dependency is much weaker than theoretically anticipated. A potential problem in this analysis are the rather large systematic uncertainties of the KIC temperatures. We therefore redid the analysis for the 624 stars in our sample for which well-calibrated effective temperatures from Sloan Digital Sky Survey photometry \citep{pinsonneault12} are available but found no significant differences in the fit. Nonetheless, our sample is dominated by red giants, which naturally cover only a relatively small range in effective temperature (a few hundred K at a given \lg , with typical uncertainties of 150\,K for individual stars). As a result it is difficult to define any temperature dependency in our sample of \textit{Kepler} stars. However, the dispersion along the fit is with about 9\% rather small so that we can expect no other properties of the star to play a significant role. For completeness we also fit $\tau_\mathrm{eff}$ as a function of \num\ and find it to approximately scale as $\tau_\mathrm{eff} \propto \nu_\mathrm{max}^{-0.89}$, which is in perfect agreement to what was found by others \citep[e.g.,][]{mat11}.

For comparison we show a power law fit to the theoretical values of $\tau_\mathrm{eff}$ (dashed line) presented by \citet{mat11} and obtained from 3D simulations of convective red-giant atmospheres \citep{tra13}. They do not provide the effective temperatures of their models, which is why we restricted the fit to the surface gravity. Interestingly, the exponent is exactly what we obtain from the observations, only the absolute values are smaller by about 60\%. This shift in absolute value is likely caused by the fact that we fit the observed granulation signal with two individual components, whereas the theoretical spectra are fitted only with one component as the kink around $\nu_\mathrm{max}/2$ is (due to presumably missing physics in the models) not present in the theoretical spectra. If we would fit the observations with only one component, as we have done for, e.g., KIC 7949599 in our background model comparison, the resulting $\tau_\mathrm{eff}$ is with $\sim$$1.17\cdot 10^4$\,s about 60\% smaller than the $\sim$$1.99\cdot 10^4$\,s that we determine for a two-component fit. 

\cite{sam13} computed granulation timescales and intensity fluctuations for a grid of 3D hydrodynamical models based on an approach described by \citet{sam13a}. Their grid also includes three K-type and three F-type dwarf models. While the K-type models are not covered by our observations, the F-type models are known to systematically overestimate the granulation amplitudes \citep[see also][]{ludwig09}. We therefore exclude them from the further analysis and plot the remaining models in Fig.\,\ref{fig:efoldtime}. From a power-law fit we find the characteristic timescales of the models to be almost perfectly reproduced as $\tau_\mathrm{eff} \propto g^{-0.86} T^{-1.2}$ (the residuals scatter by less than 3\%). The correlation with $g$ is fully compatible with the observed one (Eq.\,\ref{eq:taueff}) as well as with the theoretically expected. The temperature dependency, however, is much stronger as for the observations but weaker than originally anticipated. This might be because the models cover a different range in mass (1.4 - 4\,M\sun\ compared to $\sim$0.7 - 2.5\,M\sun\ and therefore different temperatures) on the RGB and a larger range in effective temperature ($\sim$1200\,K compared to a few hundred K) on the main sequences than the \textit{Kepler} sample and the resulting scaling might therefore be of different sensitivity to the effective temperature. To solve this discrepancy is, however, beyond the scope of this paper. As for the \citet{mat11} simulations the absolute values of model $\tau_\mathrm{eff}$ underestimate the observations (by about 40\,\%), which is again likely due to the missing kink in the theoretical spectra. Hence, apart from the missing kink and a minor issue with the temperature dependency (for the observed sample, changing the exponent from -0.4 to -1.2 changes $\tau_\mathrm{eff}$ by a few percent at most), the agreement between the observations and the model predictions are surprisingly good.

\subsection{Intensity fluctuation, $A_\mathrm{gran}$}
The rms amplitudes we measured for the two granulation components represent the rms intensity fluctuations in the time series that originate from two  physical processes on the stellar surface and that are both likely associated to granulation. It is however not relevant for our purpose to identify the exact physical process that causes each part of the signal. We are more interested in the total background signal (or so to say a proxy for the total energy) locally around \num\ and therefore have to merge the two individual amplitudes to a single value that can be compared to other observations or model predictions. This is straightforward and we define the total bolometric intensity fluctuation due to granulation as $A_\mathrm{gran}^2 = C_\mathrm{bol}^2 (a_1^2 + a_2^2)$, where $C_\mathrm{bol}$ is a bolometric correction that scales for the \textit{Kepler} bandpass as $C_\mathrm{bol} = (T_\mathrm{eff}/T_\mathrm{0})^{\alpha}$, where $T_\mathrm{0} = 5934$K and $\alpha = 0.8$ \citep{ballot11,mic09}. The estimated uncertainties of $\pm$250\,K in $T_\mathrm{eff}$ typically add about 1-2\% uncertainty to $A_\mathrm{gran}$, which are then about 7\% on average.

In a first step we fit $A_\mathrm{gran}$ as a function of the peak frequency and find it to approximately scale as $A_\mathrm{gran} \propto \nu_\mathrm{max}^{-0.56}$. This roughly corresponds to $P_\mathrm{gran} \propto \nu_\mathrm{max}^{-2.1}$ and is therefore different from what was found by, for example, \citet{mat11}, who derived an exponent of 1.89.

In the bottom panel of Fig.\,\ref{fig:efoldtime} we show $A_\mathrm{gran}$ as a function of \lg\ for our total sample of stars. A power law fit indicates that $A_\mathrm{gran} \propto g^{-0.55}$, which is in agreement to what one would expect. However, there is some structure in the correlation indicating a more complex scaling relation and there are a number of stars that fall significantly below the fit. We therefore follow the same approach as in Sec.\,\ref{Sec:GranAmp} and add a mass term to the power law, which again considerably improves the fit (in terms of global evidence of the fit and the rms scatter of the residuals). The best fit coefficients and their uncertainties are listed in Tab.\,\ref{tab:fitpar}. 

We could not compare our measurements to the models presented in \cite{mat11} as the authors did not provide estimates for the intensity fluctuations. \cite{sam13}, however, do and we plot their values in the bottom panel of Fig.\,\ref{fig:efoldtime} for comparison. Whereas the models approximately reproduce the observations on the red-giant branch, they are different for main-sequence stars. For a given \lg , the models cover a much larger range in $A_\mathrm{gran}$ than the observations. This might be because  $A_\mathrm{gran}$ is supposed to be also sensitive to the effective temperature and the main-sequence models cover a wider range in \teff . To test this we add a temperature term to our power law and find the intensity fluctuations to scale as $A_\mathrm{gran} \propto g^{-0.56}M^{-0.24}T^{0.05}$. Even though we find a quite weak temperature dependency, the global evidence of the fit is much better than for the fit without the temperature term indicating again an improvement in the scaling relation. For the sake of simplicity we ignore the temperature dependency and the granulation amplitude can then be approximated to scale as
\begin{equation} \label{eq:agran}
A_\mathrm{gran} \propto (g^2M)^{-s}, 
\end{equation}
where $s \simeq \sfrac{1}{4}$. We note that it might be more convenient to express this scaling as $A_\mathrm{gran} \propto R/M^{3/4}$ (since $g$ already depends on $M$) but prefer it the way given above as $g$ is the parameter that can be determined best from the observations and the mass dependency is only weak. Eq.\,\ref{eq:agran} is, however, incompatible with the classical scaling relation \citep[e.g.,][]{kjeldsen11} $A_\mathrm{gran} \propto (\nu_\mathrm{max} M)^{-1/2} T^{3/4}$, which translates into $A_\mathrm{gran} \propto T(gM)^{-\sfrac{1}{2}}$. 

As for the characteristic timescales, we fit our observational scaling relation to the model predictions from
\cite{sam13} and find them to approximately scale as $A_\mathrm{gran} \propto g^{-0.88}M^{-0.61}T^{6.0}$. Apart from the different exponents for $g$ and $M$, the strong dependancy on $T$ with such a high power explains the large dispersion in Fig.\,\ref{fig:efoldtime} but is not compatible to what we find for the observations. Part of the discrepancy should come from the differences of overall stellar properties between our sample and the synthetic models.

In Fig.\,\ref{fig:efoldtime} we also indicate the relation between the 8-hr flicker amplitude and \lg\ from \citet{bast13}, who demonstrated that the rms scatter of the \textit{Kepler} time series (after applying a 8 hour high-pass filter) can be used to accurately estimate the surface gravity of stars that show a surface granulation signal. Interestingly, their flicker amplitude almost perfectly resembles our $A_\mathrm{gran}$ scaling for stars with a \lg\ between about 3 to 3.5, indicating that they indeed measure the intensity fluctuations due to granulation. For stars with smaller or larger \lg , however, the flicker amplitude represents only a part of the granulation amplitude. We note that this method is not applicable for many of the red giants because the fixed 8 hour filter suppresses the granulation (and pulsation) signal in stars with \num\ $\lesssim$ 35\mh\ (i.e., with periods longer than 8 hours).

\begin{figure}[t]
	\begin{center}
	\includegraphics[width=0.5\textwidth]{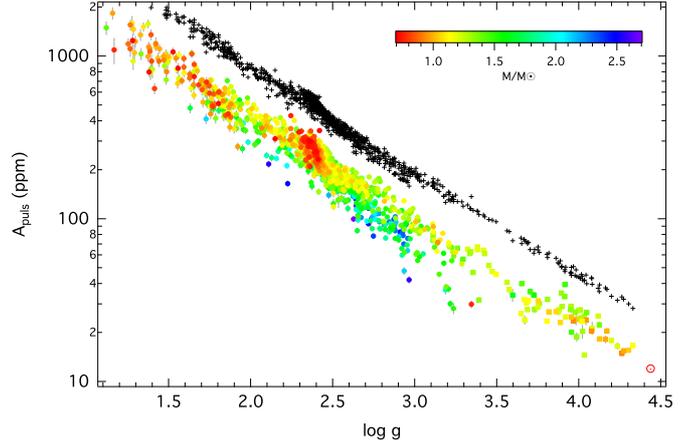}
	\caption{Total pulsation amplitude as a function of \lg\ for our sample of LC (circles) and SC (squares) stars, with the colour encoding stellar mass. Black crosses indicate pulsation amplitudes predicted from a power law fit of the form $A_\mathrm{puls} \propto (g^2 M)^{-s}$ (shifted by a factor of 2 for better visibility). } 
	\label{fig:apuls} 
	\end{center} 
\end{figure}

\subsection{Pulsation amplitudes}	\label{sec:apuls}
Oscillation amplitudes are a difficult to measure and model asteroseismic quantity. From a theoretical point of view, the amplitude of a mode is determined by the competing driving and damping mechanism, involving rather complex physics \citep[e.g.,][]{houdek99,sam07}. In solar-type oscillating stars, the outer convective layer is believed to drive modes with a resonant frequency $v_\mathrm{CV}/H_p$, where $v_\mathrm{CV}$ is the convection velocity, and a mode energy that is roughly equal to the kinetic energy of a single convection eddy. The underlying physics is still not properly understood but a number of scaling relations aiming to predict mode amplitudes by scaling from the Sun's values have been derived and discussed \citep[see, e.g.,][and references therein]{corsaro13}. Testing such scaling relations with observational data is vital for a better understanding of stellar oscillations but not only their physical interpretation is still a matter of debate, also the actual measurements are far from being straightforward to perform.

Ideally, it would require to fit a sequence of Lorentzian profiles on top of an appropriate background model to determine the individual mode amplitudes. This is already a challenging task for individual stars with good SNR and frequency resolution \citep[e.g.,][]{gru09} and currently practically impossible for a large sample of stars. In practice, the power density spectra are heavily smoothed, corrected for the background signal and converted to amplitude per oscillation mode. This requires not only a good knowledge of the granulation background (which is often not the case, see Sec.\,\ref{BGmodel}) but is also based on the assumption that modes of different degree having a different spatial response are excited to the same intrinsic amplitude \citep[e.g.,][]{kjeldsen08}, which is also often not the case (see next. Sec.). Furthermore, sampling effects as described in Sec.\,\ref{AmpDamp}, which potentially introduce significant systematic uncertainties are frequently not accounted for. Alternatively, we use what we call the total pulsation amplitude $A_\mathrm{puls}$, which is a direct output of our power spectra modelling and which represents the sum of the amplitudes of \textit{all} excited modes (even those that are not detectable individually) and is therefore a good measure for the total intensity fluctuations due to oscillations that we observe on the stellar surface. 

From theoretical considerations it is assumed that the bolometric mode amplitude scales as $(L/M)^p (T_\mathrm{eff})^{-t}$, where it is not yet fully clear whether or not it scales with $L$ and $M$ to the same power and what are the specific values of the exponents \citep[e.g.,][]{ste11, hub11, corsaro13}. For the vast majority of the rather faint stars observed with \textit{Kepler}, however, the luminosity is simply unknown (or poorly determined via the seismic radius and effective temperature) and it is therefore difficult to test such a scaling relation. Instead we prefer to adopt the effective temperature and surface gravity as independent variables since $L/M \propto T_\mathrm{eff}^4/g$ (given $g \propto M/R^2$ and $L \propto R^2 T_\mathrm{eff}^4$). \citet{hub11} and \citet{ste11} suggested independent exponents for $L$ and $M$, which means that we have to add a mass term to our scaling relation. This can also be seen from Fig.\,\ref{fig:apuls}, where we plot $A_\mathrm{puls}$ as a function of \lg , with the stellar mass colour-coded. Clearly, mass affects the pulsation amplitudes so that low-mass stars tend to have larger amplitudes than high-mass stars at a given \lg . A power law fit indicates that the total (bolometric) pulsation amplitude roughly scales as $A_\mathrm{puls} \propto g^{-0.66} M^{-0.35} T^{0.04}$. As for the granulation amplitude, \textit{MultiNest} found the exponent of the temperature term to be close to zero, which means that the pulsation amplitude of a star can be sufficiently well explained by the stars mass and surface gravity (i.e., mass and radius) and $T_\mathrm{eff}$ has no (or only marginal) effect. Therefore we can approximate the pulsation amplitude to scale as
\begin{equation} \label{Apuls1}
A_\mathrm{puls} \propto (g^{2} M)^{-s}, 
\end{equation}
where $s \simeq \sfrac{1}{3}$. Substituting $g$ by $M/R^2$ and $R^2$ by $L/T^4$, this translates to 
\begin{equation} \label{Apuls2}
A_\mathrm{puls} \propto R^{4/3}/M \propto (L^{2} /M^{3})^s T^{-8s}, 
\end{equation}
showing that $L$ and $M$ indeed have different exponents. To test the statistical significance of this we also fitted a power law $A_\mathrm{puls} \propto g^s T^{-t}$ (i.e. without an additional mass term and therefore $L/M$ with a single exponent) and find the odds ratio between the two model's global evidence clearly in favour for the scaling relation including a mass term.

In principle Eqs.\,\ref{Apuls1} and \ref{Apuls2} are equivalent but from an observational point of view, Eq.\,\ref{Apuls1} should be preferred as it uses easier to determine parameters.
 
The relative scatter around the fit of about 13\% leaves not much space for other properties of the star to significantly affect the pulsation amplitudes. However, there are a number of stars that have substantially lower total amplitudes than what one expect from their M and \lg . This could be caused by, for example, binaries or stellar activity damping the oscillations \citep[see, e.g.,][]{hub11}.

From Fig\,\ref{fig:apuls} it appears the low-mass stars are overabundant on the upper giant branch (\lg\ $\lesssim$ 2.0) compared to the stars below the red clump (at \lg\ $\sim$ 2.3). This is presumably due to the mass loss that stars undergo when evolving up the giant branch \citep[see, e.g.,][]{mig11}.

\section{Granulation versus pulsation amplitudes}		\label{sec:granpuls}

Solar-like oscillations are excited by convection, which is also the process responsible for granulation. The basic assumption is that the power of the velocity fluctuations due to p-mode oscillations scales with stellar parameters in the same way as the power of the velocity fluctuations due to granulation. This is roughly supported by solar observations indicating that both the kinetic energy of an oscillation mode and of a single granule are about $10^{27}$\,ergs \citep{Korzennik12}. Many authors \citep[e.g.,][]{kjeldsen11} have previously argued that amplitudes of solar-like oscillations should scale in proportion to fluctuations due to granulation. This is also supported by our findings, where we show that the pulsation amplitudes scale to nearly the same power of $g$ and $M$ as the granulation intensity fluctuations. It is therefore obvious to correlate these two parameters.

\begin{figure}[t]
	\begin{center}
	\includegraphics[width=0.5\textwidth]{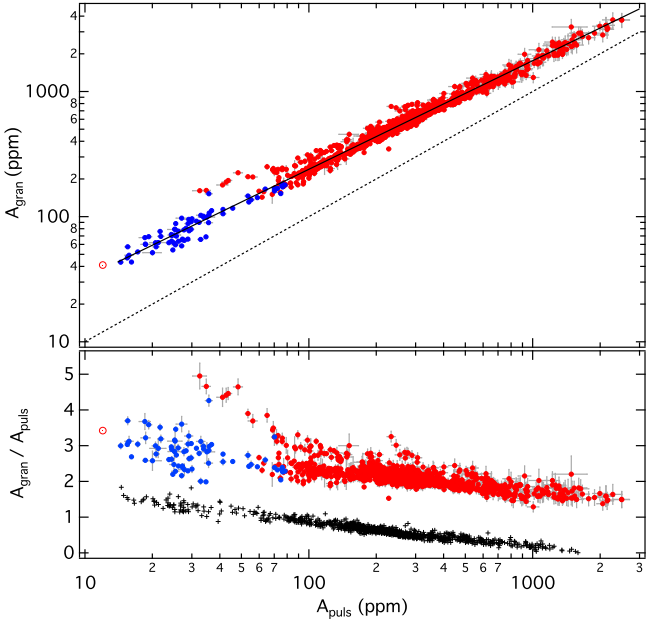}
	\caption{Granulation intensity fluctuations $A_\mathrm{gran}$ as a function of the total pulsation amplitude $A_\mathrm{puls}$. Symbol colours are the same as in Fig.\,\ref{fig:ab_numax}. The dotted and solid line indicate unity and a power law fit, respectively. Black crosses indicate the ratio $A_\mathrm{gran}/A_\mathrm{puls}$ predicted from a power law fit $A_\mathrm{gran} \propto A_\mathrm{puls}^s g^t M^u$ (shifted by -1.5 in y-direction for better visibility). Note that the fit does not include the weak dipole mode stars.} 
	\label{fig:ampptot} 
	\end{center} 
\end{figure}

This is done in Fig.\,\ref{fig:ampptot}, where we show the amplitude of the granulation intensity fluctuations as a function of the total pulsation amplitude. Even though there is a tight relation, there is no equality nor is it a linear relation. A power law fit indicates that $A_\mathrm{gran} \propto A_\mathrm{puls}^{0.86}$. The bottom panel of Fig.\,\ref{fig:ampptot} shows that while the granulation amplitude is about 3 to 4 times larger than the pulsation amplitude for stars on the main sequence, the ratio drops down to about 1.5 for stars high up on the giant branch. 

The fact that there is no linear relation between the granulation and pulsation amplitude is somewhat surprising as one would intuitively expect that a more vigorous convection causes a stronger granulation signal in the same way as larger pulsation amplitudes. On the other hand we have to keep in mind that the amplitude of solar-like oscillations depends on both the excitation (i.e., the amount of energy that is provided by convection) as well as the damping rate (i.e., the mode lifetime). \citet{kjeldsen11} postulated that the squared p-mode amplitude in velocity scales with the granulation power at \num\ times the mode lifetime. Efforts to define a scaling relations for the mode lifetime have been made \citep{chaplin09, baudin11, cor12} but given the fact that measuring mode lifetimes is still quite challenging no consensus has been found so far. It is, however, well established that the mode lifetime is a function of temperature.

\begin{figure}[b]
	\begin{center}
	\includegraphics[width=0.5\textwidth]{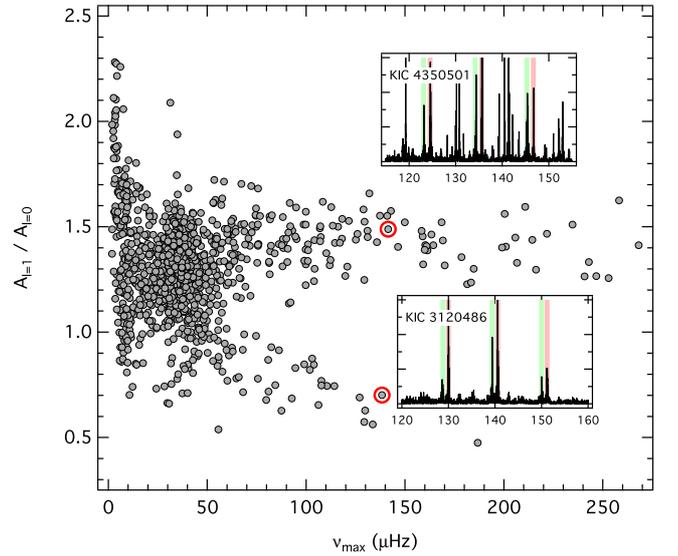}
	\caption{Ratio between the average amplitude of the central two $l$ = 1 and the central three $l$ = 0 modes, showing a group of stars with atypically small dipole modes ($A_{l=1}/A_{l=0} < 1$). Inserts show the power density spectum of two with red open circles marked stars, where coloured bars label $l$ = 0 (lightred) and 2 (lightgreen) modes.} 
	\label{fig:ampratio} 
	\end{center} 
\end{figure}

To account for this we add a temperature term and find that the granulation intensity fluctuations approximately scale as $A_\mathrm{gran} \propto A_\mathrm{puls}^{0.84}T^{0.24}$. Even though there is a relatively good correlation between $A_\mathrm{gran}/A_\mathrm{puls}^s$ and $T$ (see Tab.\,\ref{tab:fitpar}), including the temperature significantly impairs the fit. This is reflected by the increasing rms scatter of the residuals from $\sim$8 to $\sim$10.3\% as well as a much lower global model evidence. A much better description of the correlation between the granulation and pulsation amplitude can be given if we include the surface gravity and mass in the fit. We find the granulation amplitude to scale as $A_\mathrm{gran} \propto A_\mathrm{puls}^{0.63}\, g^{-0.15} M^{-0.06}$, which is consistent with what one can expect when comparing the individual scalings for the granulation and pulsation amplitudes.

\subsection{Weak dipole modes}
Fig.\,\ref{fig:ampptot} shows an interesting feature with a group of stars deviating from the general trend. From a pulsation amplitude of about 200\,ppm downwards there exists a group of stars that seem to have substantially larger granulation amplitudes than expected. A closer look on the individual power spectra, however, reveals that it is the total pulsation amplitude that is smaller than it should be for ``normal'' stars. This is because the dipole modes are much weaker (or even totally suppressed) than the surrounding $l$ = 0 and 2 modes, which is reflected in the integrated amplitude as there is some power ``missing''. The origin of this phenomena is unknown but \citet{mos12a} and only recently \cite{garcia13} found that it is not correlated with, e.g., the temperature or chemical composition of these stars, and \citet{mosser13} argue that it likely results from a very efficient coupling between pressure and gravity waves. 

To check whether this effect is also reflected in the granulation signal that excites the modes, we need to estimate how much pulsation power is missing. We therefore compute the average amplitude of the central two $l$ = 1 modes and the central three $l$ = 0 and 2 modes. Following \citet{mos12a} the individual mode amplitudes are given as $A^2_{l}(\nu) = \delta\nu \sum_{\delta^{-}}^{\delta^{+}} [P(\nu)-B(\nu)]$, where $\delta\nu$ is the bin width of independent frequency bins with the power density $P(\nu)$ and background level $B(\nu)$. In the case of $l$ = 0 and 2 modes, the spectrum is evaluated between $\delta^{\pm} = \nu_m \pm \delta\nu_{02}/2$ (with $\nu_m$ and $\delta\nu_{02}$ being the mode frequency and the small frequency separation, respectively) and in the case of dipole modes, from $\delta^{-} = \nu_0+\delta\nu_{02}/2$ to $\delta^{+} = \nu_2-\delta\nu_{02}/2$ (with $\nu_0$ and $\nu_2$ being the adjacent $l$ = 0 and 2 mode frequencies). The amplitude ratio $A_1/A_0$ is given in Fig.\,\ref{fig:ampratio} and clearly shows that stars with  \num\ $\gtrsim$ 60\mh\ split into two groups\footnote{$A_2/A_0$ does not show this effect, which is why we do not plot it in Fig.\,\ref{fig:ampratio}. Its average value is about 0.85.}. While normal stars have an amplitude ratio of $\sim$1.4 on average, $A_1/A_0$ reduces to  $\sim$1.0 to 0.5 (depending on \num ) in weak dipole-mode stars. Assuming that the total pulsation amplitude is roughly proportional to $A_0 (1 + A_1/A_0 + A_2/A_0)$ we can estimate the $A_\mathrm{puls}$ is underestimated by about 15 to 40\% compared to normal stars. Adding this ``missing''  pulsation amplitude in Fig.\,\ref{fig:ampptot} would shift the weak dipole-mode stars into (or at least close to) the population of normal stars. From this we conclude that the granulation signal acts normal in such stars confirming the early result based on KIC 8561221 in which the background was also found to be normal \citep{garcia13}. This needs, however further work, which is beyond the scope of this paper. We therefore excluded these stars from our analysis.

\section{Testing the scaling relations on the Sun}
A good possibility for an independent test of the above deduced scaling relations is given by the Sun. We use a 1-year time series from the green channel of the SOHO/VIRGO data \citep{fro97} obtained during the solar activity minimum before Cycle 23 and fitted our model F to the corresponding power density spectrum. The resulting best-fit parameter are listed in Tab.\,\ref{tab:Sun}, where $A_\mathrm{gran,\odot}$ and $A_\mathrm{puls,\odot}$ are bolometrically corrected according to \citet{mic09}. From this analysis it becomes again obvious that the specific treatment of the granulation background has an important impact on the determination of \num . Whereas \citet{kal10b} fitted the same model to the same data, they did not arrive at the same \num . This is presumably due to the fact that we now do also account for the sampling effects described in Sec.\,\ref{AmpDamp}, which redistributes the power in the power excess and shifts \num\ from 3120$\pm$5 to 3140$\pm$4\mh .

In Tab.\,\ref{tab:Sun} we also give the values that result from our scaling relations and find them in good agreement with the actual measurements. Note that the predicted amplitudes $a_{1,\odot}$ and $a_{2,\odot}$ are given for the SOHO/VIRGO bandpass (centred on 500\,nm), where we assume a simple linear transformation from the \textit{Kepler} bandpass (centred on 664\,nm).

\begin{table}[t]
\begin{center}
\caption{Solar reference values and the corresponding parameter predicted from scaling realtions. The predicted amplitudes $a_{1,\odot}$ and $a_{2,\odot}$ are multiplied by 1.33 to account for the different passbands of \textit{Kepler} and SOHO/VIRGO. $A_\mathrm{gran,\odot}$ and $A_\mathrm{puls,\odot}$ are bolometric values. The units are the same as in Tab.\,\ref{tab:BGmodels} and \ref{tab:fitpar}.
\label{tab:Sun}}
\begin{tabular}{c|r|r}
\hline
&solar values&scaling relation\\
\hline
\noalign{\smallskip}
$P_{g,\odot}$&0.25(1)&\\
$\nu_\mathrm{max,\odot}$&3140(4)&\\
$\sigma_{\odot}$&361(4)&\\
\noalign{\smallskip}
\hline
\noalign{\smallskip}
$a_{1,\odot}$&36.5(2)&35.5(7)\\
$b_{1,\odot}$&758(8)&782(11)\\
$a_{2,\odot}$&35.9(2)&35.5(7)\\
$b_{2,\odot}$&2468(9)&2787(43)\\
\noalign{\smallskip}
\hline
\noalign{\smallskip}
$\tau_\mathrm{eff,\odot}$&375(4)&340(12)\\
$A_\mathrm{gran,\odot}$&41.0(2)&38(2)\\
$A_\mathrm{puls,\odot}$&12.01(6)&11.4(6)\\
\noalign{\smallskip}
\hline
\end{tabular}
\end{center}
\end{table}

\section{Summary}

In this work we investigated the granulation background spectrum of a large and homogeneous sample of 1364 stars observed by \textit{Kepler}. The sample includes stars from the main sequence, the sub-giant branch to stars on the ascending giant branch, in the red clump, and on the asymptotic giant branch. Or with other words, stars with a mass ranging from about 0.7 to 2.5\,M\sun\ that are cooler than the red border of the classical instability strip, that cover a large fraction of the parameter space for which we can expect convective surface layers that exhibit solar-like oscillations and granulation. We used \textit{Kepler} light curves that span between 90 and 360 days for main-sequence stars and sub-giants and about 1140 days for red giants, corrected them for various instrumental effects and analysed the overall structure of the resulting power density spectra. From this study we find that:
\begin{itemize}
  \item The depression in power at about \num/2 that is know from the Sun and a few other main-sequence stars is a common feature in all stars of our sample. A plausible explanation for this feature is the presence of a second background component close to the pulsation power excess. The fact that the position of the depression relative to the power excess is roughly constant for all stars indicates that the characteristic timescales and amplitudes of the two components have a fixed ratio. 
\item Using two (instead of one) background components significantly improves the estimate of the background underneath the oscillation signal. A single component is only sufficient if the instrumental white noise dominates the background (or is at least not negligible as for many of the \textit{Kepler} targets) hiding the specific shape of the granulation signal. 
  \item In a probabilistic comparison of various functional forms of background models a super-Lorentzian function with a free (but close to 4) exponent turned out to reproduce the observed signal best for many stars. The data do, however, not (yet) provide enough evidence for such a model and a model with an exponent fixed to 4 (which represents a significant simplification for the fitting) can be considered as equally good.
  \item The specific choice of the background model influences the determination of \num . An inappropriate background model can redistribute the power in the pulsation power excess and systematically shift its centre. We find such shifts in a local (around \num ) as well as a global analysis of the power spectrum and note that the systematics are largest between one and two-component models. We confirmed these findings using simulated power spectra.
\end{itemize} 

 We then used a Bayesian inference tool to determine the granulation and global oscillation parameters and their uncertainties for our sample of stars. Compared to the previous analysis of \citet{mat11} we extended the sample towards sub-giant branch and main-sequence stars and used time series for the red giants that are about three times as long. On the other hand, M giants that show solar-like oscillations \citep{mos13a} are excluded from our analysis as their granulation signal has such long periods that we cannot derive precise parameters even with the long \textit{Kepler} time series. The total sample now covers more than 3 order of magnitudes in \num . From the analysis of the resulting parameters we conclude that:

\begin{itemize}
\item There are tight relations between all granulation parameters and \num . Whereas the characteristic granulation frequencies scale almost linearly with \num , the granulation amplitudes approximately scale as $\nu_\mathrm{max}^{-0.6}$. We also confirmed a significant mass dependency of the latter. 
\item The defined scaling relations allow to estimate the overall shape of the granulation signal of any solar-type oscillator to within about 15\%, and might therefore serve as a good starting point for future large sample studies.
\item In order to allow a comparison with previous measurements and model estimates we also computed effective timescales and total intensity fluctuations. We established that both parameters are predominantly determined by gravity on the stellar surface. They were found to approximately scale as $\tau_\mathrm{eff} \propto g^{-0.85}T^{-0.4}$ and $A_\mathrm{gran} \propto (g^2M)^{-1/4} \propto R/M^{3/4}$. From the rather small rms scatter of the residuals ($<$10\% for both parameters) we conclude that no other property of the star plays a significant role.  
\item While the theoretical predictions of the characteristic timescales are generally in good agreement with the observation, some discrepancies remain for the intensity fluctuations.
\item There is a statistically significant and surprisingly simple scaling relation for the total pulsation amplitude, which turned out to approximately scale as $A_\mathrm{puls} \propto (g^2M)^{-1/3} \propto R^{4/3}/M$. This implicitly verifies a separate mass and luminosity but no (additional) temperature dependence of the mode amplitudes. Our sample certainly includes stars with diminished pulsation amplitudes due to, e.g., an increased activity level or binary companions. The small rms scatter of the residuals of about 14\% therefore indicates that the unperturbed pulsation amplitudes are well approximated by the above scaling relation.
\item The granulation timescale and amplitude as well as the pulsation amplitude can be sufficiently described by the mass and surface gravity of a star and the effective temperature has no (or only marginal) additional effect on these parameters.
\item The granulation signal in weak dipole-mode stars is at first order indistinguishable from that of normal stars.
\end{itemize}

\appendix
\section{Power density conversion} \label{PDSconversion}

Measuring amplitudes of a quasi-stochastic signal (i.e., solar-like oscillations but also the granulation signal) is not straight forward. What we actually measure in a power spectrum is, e.g., the height of an oscillation mode (i.e., the amplitude of the limit spectrum). This does not only scale with the intrinsic mode amplitude and line width but also with the time baseline of the observations. It is therefore convenient to convert the spectral power $\mathcal{P}_{PS}$ to spectral power density $\mathcal{P}_{PDS}$, to become independent of the baseline. There are several approaches to do so. A frequently used method \citep[e.g.,][]{Appourchaux08} is to search for a conversion factor $\zeta$ so that 
\begin{equation}\label{eq:par}
\sigma^2 = \zeta \int_{0}^{\nu_{nq}} \mathcal{P}_{PS}(\nu)\,d\nu 
\end{equation}
is satisfied (also known as Parseval's theorem), where $\sigma^2$ and $\nu_{nq}$ are the variance and Nyqvist frequency\footnote{In the barycentric reference frame, the median sampling for \textit{Kepler} LC and SC time series is about 29.38 minutes and 58.85 seconds, respectively, from which we determine $\nu_{nq}$ to about 283.6 and 8496\mh .} of the time series, respectively. The spectral power density is then given as $\mathcal{P}_{PDS} = \zeta \mathcal{P}_{PS}$. Strictly speaking, this relation is only valid for evenly spaced and continuous data sets. However, introducing gaps in a continuous time series, that contains for example only Gaussian noise, will not change its variance but it will change the integral of the corresponding power spectrum and therefore the conversion factor $\zeta$. This is because the gaps in the time series produce alias peaks in the spectral window function that provide additional power to the spectrum, which is not present in the time domain. To account for this, sort of a filling factor could be introduced in Eq.\,\ref{eq:par} but it is more convenient to use the actual spectral window function $\mathcal{S}_{PS}$. We define the conversion factor between spectral power and spectral power density as, 
\begin{equation}\label{eq:zeta}
\zeta = \left (\int_{-\nu_{nq}}^{+\nu_{nq}} \mathcal{S}_{PS}(\nu)\,d\nu \right)^{-1}.
\end{equation}
That said, we note that for most of the \textit{Kepler} time series the conversion factors resulting from Eqs.\,\ref{eq:par} and \ref{eq:zeta} are almost identical because the spectral window functions of the high-duty-cycle time series are close to a Dirac function. However, for stars with large gaps (e.g., due to missing quarters) the difference can be non-negligible.

\begin{acknowledgements}
The authors gratefully acknowledge the \textit{Kepler Science Team} and all those who have contributed to making the \textit{Kepler} mission possible. Funding for the \textit{Kepler Discovery mission} is provided by NASA's Science Mission Directorate. TK and JDR are supported by the FWO-Flanders under project O6260 - G.0728.11. TK acknowledges financial support from the Austrian Science Fund (FWF P23608). SH acknowledges support from the European Research Council under the European Community’s Seventh Framework Programme (FP7/2007-2013) / ERC grant agreement no 338251 (StellarAges) as well as from the Deutsche Forschungsgemeinschaft (DFG) under grant SFB g63/1, ``Astrophysical flow instabilities and turbulence''. BM acknowledges financial support from the `Programme National de Physique Stellaire' (PNPS, INSU, France) of CNRS/INSU and from the ANR program IDEE `Interaction Des \'Etoiles et des Exoplan\`etes' (Agence Nationale de la Recherche, France). This work partially used data analysed under the NASA grant NNX12AE17G and under the European Community Seventh Framework Program grant (FP7/2007-2013)/ERC grant agreement n. PROSPERITY. SOHO is a mission of international collaboration between ESA and NASA.

\end{acknowledgements}

\bibliographystyle{aa}
\bibliography{24313}

\begin{thebibliography}{79}
\expandafter\ifx\csname natexlab\endcsname\relax\def\natexlab#1{#1}\fi

\bibitem[{{Aigrain} {et~al.}(2004){Aigrain}, {Favata}, \&
  {Gilmore}}]{aigrain04}
{Aigrain}, S., {Favata}, F., \& {Gilmore}, G. 2004, A\&A, 414, 1139

\bibitem[{{Andersen} {et~al.}(1998){Andersen}, {Leifsen}, {Appourchaux},
  {Frohlich}, {Hoeksema}, \& {Toutain}}]{and98}
{Andersen}, B., {Leifsen}, T., {Appourchaux}, T., {et~al.} 1998, in ESA Special
  Publication, Vol. 418, Structure and Dynamics of the Interior of the Sun and
  Sun-like Stars, ed. S.~{Korzennik}, 893

\bibitem[{{Antoci} {et~al.}(2011){Antoci}, {Handler}, {Campante}, {Thygesen},
  {Moya}, {Kallinger}, {Stello}, {Grigahc{\`e}ne}, {Kjeldsen}, {Bedding},
  {L{\"u}ftinger}, {Christensen-Dalsgaard}, {Catanzaro}, {Frasca}, {De Cat},
  {Uytterhoeven}, {Bruntt}, {Houdek}, {Kurtz}, {Lenz}, {Kaiser}, {van Cleve},
  {Allen}, \& {Clarke}}]{antoci11}
{Antoci}, V., {Handler}, G., {Campante}, T.~L., {et~al.} 2011, Nature, 477, 570

\bibitem[{{Appourchaux} {et~al.}(2008){Appourchaux}, {Michel}, {Auvergne},
  {Baglin}, {Toutain}, {Baudin}, {Benomar}, {Chaplin}, {Deheuvels}, {Samadi},
  {Verner}, {Boumier}, {Garc{\'{\i}}a}, {Mosser}, {Hulot}, {Ballot}, {Barban},
  {Elsworth}, {Jim{\'e}nez-Reyes}, {Kjeldsen}, {R{\'e}gulo}, \&
  {Roxburgh}}]{Appourchaux08}
{Appourchaux}, T., {Michel}, E., {Auvergne}, M., {et~al.} 2008, A\&A, 488, 705

\bibitem[{{Baglin} {et~al.}(2006){Baglin}, {Auvergne}, {Barge}, {Deleuil},
  {Catala}, {Michel}, {Weiss}, \& {The COROT Team}}]{bag06}
{Baglin}, A., {Auvergne}, M., {Barge}, P., {et~al.} 2006, in ESA Special
  Publication, Vol. 1306, ESA Special Publication, ed. {M.~Fridlund, A.~Baglin,
  J.~Lochard, \& L.~Conroy}, 33

\bibitem[{{Ballot} {et~al.}(2011){Ballot}, {Barban}, \& {van't
  Veer-Menneret}}]{ballot11}
{Ballot}, J., {Barban}, C., \& {van't Veer-Menneret}, C. 2011, A\&A, 531, A124

\bibitem[{{Bastien} {et~al.}(2013){Bastien}, {Stassun}, {Basri}, \&
  {Pepper}}]{bast13}
{Bastien}, F.~A., {Stassun}, K.~G., {Basri}, G., \& {Pepper}, J. 2013, Nature,
  500, 427

\bibitem[{{Baudin} {et~al.}(2011){Baudin}, {Barban}, {Belkacem}, {Hekker},
  {Morel}, {Samadi}, {Benomar}, {Goupil}, {Carrier}, {Ballot}, {Deheuvels}, {De
  Ridder}, {Hatzes}, {Kallinger}, \& {Weiss}}]{baudin11}
{Baudin}, F., {Barban}, C., {Belkacem}, K., {et~al.} 2011, A\&A, 529, A84

\bibitem[{{Beck} {et~al.}(2012){Beck}, {Montalban}, {Kallinger}, {De Ridder},
  {Aerts}, {Garc{\'{\i}}a}, {Hekker}, {Dupret}, {Mosser}, {Eggenberger},
  {Stello}, {Elsworth}, {Frandsen}, {Carrier}, {Hillen}, {Gruberbauer},
  {Christensen-Dalsgaard}, {Miglio}, {Valentini}, {Bedding}, {Kjeldsen},
  {Girouard}, {Hall}, \& {Ibrahim}}]{beck11b}
{Beck}, P.~G., {Montalban}, J., {Kallinger}, T., {et~al.} 2012, Nature, 481, 55

\bibitem[{{Bedding} {et~al.}(2011){Bedding}, {Mosser}, {Huber},
  {Montalb{\'a}n}, {Beck}, {Christensen-Dalsgaard}, {Elsworth},
  {Garc{\'{\i}}a}, {Miglio}, {Stello}, {White}, {De Ridder}, {Hekker}, {Aerts},
  {Barban}, {Belkacem}, {Broomhall}, {Brown}, {Buzasi}, {Carrier}, {Chaplin},
  {di Mauro}, {Dupret}, {Frandsen}, {Gilliland}, {Goupil}, {Jenkins},
  {Kallinger}, {Kawaler}, {Kjeldsen}, {Mathur}, {Noels}, {Aguirre}, \&
  {Ventura}}]{bed11}
{Bedding}, T.~R., {Mosser}, B., {Huber}, D., {et~al.} 2011, Nature, 471, 608

\bibitem[{{Belkacem} {et~al.}(2011){Belkacem}, {Goupil}, {Dupret}, {Samadi},
  {Baudin}, {Noels}, \& {Mosser}}]{bel11}
{Belkacem}, K., {Goupil}, M.~J., {Dupret}, M.~A., {et~al.} 2011, A\&A, 530,
  A142

\bibitem[{{Belkacem} {et~al.}(2009){Belkacem}, {Samadi}, {Goupil},
  {Lef{\`e}vre}, {Baudin}, {Deheuvels}, {Dupret}, {Appourchaux}, {Scuflaire},
  {Auvergne}, {Catala}, {Michel}, {Miglio}, {Montalban}, {Thoul}, {Talon},
  {Baglin}, \& {Noels}}]{bel09}
{Belkacem}, K., {Samadi}, R., {Goupil}, M.-J., {et~al.} 2009, Science, 324,
  1540

\bibitem[{{Borucki} {et~al.}(2010){Borucki}, {Koch}, {Basri}, {Batalha},
  {Brown}, {Caldwell}, {Caldwell}, {Christensen-Dalsgaard}, {Cochran},
  {DeVore}, {Dunham}, {Dupree}, {Gautier}, {Geary}, {Gilliland}, {Gould},
  {Howell}, {Jenkins}, {Kondo}, {Latham}, {Marcy}, {Meibom}, {Kjeldsen},
  {Lissauer}, {Monet}, {Morrison}, {Sasselov}, {Tarter}, {Boss}, {Brownlee},
  {Owen}, {Buzasi}, {Charbonneau}, {Doyle}, {Fortney}, {Ford}, {Holman},
  {Seager}, {Steffen}, {Welsh}, {Rowe}, {Anderson}, {Buchhave}, {Ciardi},
  {Walkowicz}, {Sherry}, {Horch}, {Isaacson}, {Everett}, {Fischer}, {Torres},
  {Johnson}, {Endl}, {MacQueen}, {Bryson}, {Dotson}, {Haas}, {Kolodziejczak},
  {Van Cleve}, {Chandrasekaran}, {Twicken}, {Quintana}, {Clarke}, {Allen},
  {Li}, {Wu}, {Tenenbaum}, {Verner}, {Bruhweiler}, {Barnes}, \& {Prsa}}]{bor10}
{Borucki}, W.~J., {Koch}, D., {Basri}, G., {et~al.} 2010, Science, 327, 977

\bibitem[{{Brown} {et~al.}(1991){Brown}, {Gilliland}, {Noyes}, \&
  {Ramsey}}]{bro91}
{Brown}, T.~M., {Gilliland}, R.~L., {Noyes}, R.~W., \& {Ramsey}, L.~W. 1991,
  ApJ, 368, 599

\bibitem[{{Brown} {et~al.}(2011){Brown}, {Latham}, {Everett}, \&
  {Esquerdo}}]{brown11}
{Brown}, T.~M., {Latham}, D.~W., {Everett}, M.~E., \& {Esquerdo}, G.~A. 2011,
  AJ, 142, 112

\bibitem[{{Chaplin} {et~al.}(1997){Chaplin}, {Elsworth}, {Howe}, {Isaak},
  {McLeod}, {Miller}, \& {New}}]{cha97}
{Chaplin}, W.~J., {Elsworth}, Y., {Howe}, R., {et~al.} 1997, MNRAS, 287, 51

\bibitem[{{Chaplin} {et~al.}(2009){Chaplin}, {Houdek}, {Karoff}, {Elsworth}, \&
  {New}}]{chaplin09}
{Chaplin}, W.~J., {Houdek}, G., {Karoff}, C., {Elsworth}, Y., \& {New}, R.
  2009, A\&A, 500, L21

\bibitem[{{Chaplin} {et~al.}(2011{\natexlab{a}}){Chaplin}, {Kjeldsen},
  {Bedding}, {Christensen-Dalsgaard}, {Gilliland}, {Kawaler}, {Appourchaux},
  {Elsworth}, {Garc{\'{\i}}a}, {Houdek}, {Karoff}, {Metcalfe},
  {Molenda-{\.Z}akowicz}, {Monteiro}, {Thompson}, {Verner}, {Batalha},
  {Borucki}, {Brown}, {Bryson}, {Christiansen}, {Clarke}, {Jenkins}, {Klaus},
  {Koch}, {An}, {Ballot}, {Basu}, {Benomar}, {Bonanno}, {Broomhall},
  {Campante}, {Corsaro}, {Creevey}, {Esch}, {Gai}, {Gaulme}, {Hale},
  {Handberg}, {Hekker}, {Huber}, {Mathur}, {Mosser}, {New}, {Pinsonneault},
  {Pricopi}, {Quirion}, {R{\'e}gulo}, {Roxburgh}, {Salabert}, {Stello}, \&
  {Suran}}]{cha11}
{Chaplin}, W.~J., {Kjeldsen}, H., {Bedding}, T.~R., {et~al.}
  2011{\natexlab{a}}, ApJ, 732, 54

\bibitem[{{Chaplin} {et~al.}(2011{\natexlab{b}}){Chaplin}, {Kjeldsen},
  {Christensen-Dalsgaard}, {Basu}, {Miglio}, {Appourchaux}, {Bedding},
  {Elsworth}, {Garc{\'{\i}}a}, {Gilliland}, {Girardi}, {Houdek}, {Karoff},
  {Kawaler}, {Metcalfe}, {Molenda-{\.Z}akowicz}, {Monteiro}, {Thompson},
  {Verner}, {Ballot}, {Bonanno}, {Brand{\~a}o}, {Broomhall}, {Bruntt},
  {Campante}, {Corsaro}, {Creevey}, {Do{\u g}an}, {Esch}, {Gai}, {Gaulme},
  {Hale}, {Handberg}, {Hekker}, {Huber}, {Jim{\'e}nez}, {Mathur}, {Mazumdar},
  {Mosser}, {New}, {Pinsonneault}, {Pricopi}, {Quirion}, {R{\'e}gulo},
  {Salabert}, {Serenelli}, {Silva Aguirre}, {Sousa}, {Stello}, {Stevens},
  {Suran}, {Uytterhoeven}, {White}, {Borucki}, {Brown}, {Jenkins}, {Kinemuchi},
  {Van Cleve}, \& {Klaus}}]{chap11b}
{Chaplin}, W.~J., {Kjeldsen}, H., {Christensen-Dalsgaard}, J., {et~al.}
  2011{\natexlab{b}}, Science, 332, 213

\bibitem[{{Christensen-Dalsgaard}(2002)}]{jcd02}
{Christensen-Dalsgaard}, J. 2002, Reviews of Modern Physics, 74, 1073

\bibitem[{{Corsaro} {et~al.}(2013){Corsaro}, {Fr{\"o}hlich}, {Bonanno},
  {Huber}, {Bedding}, {Benomar}, {De Ridder}, \& {Stello}}]{corsaro13}
{Corsaro}, E., {Fr{\"o}hlich}, H.-E., {Bonanno}, A., {et~al.} 2013, MNRAS, 430,
  2313

\bibitem[{{Corsaro} {et~al.}(2012){Corsaro}, {Stello}, {Huber}, {Bedding},
  {Bonanno}, {Brogaard}, {Kallinger}, {Benomar}, {White}, {Mosser}, {Basu},
  {Chaplin}, {Christensen-Dalsgaard}, {Elsworth}, {Garc{\'{\i}}a}, {Hekker},
  {Kjeldsen}, {Mathur}, {Meibom}, {Hall}, {Ibrahim}, \& {Klaus}}]{cor12}
{Corsaro}, E., {Stello}, D., {Huber}, D., {et~al.} 2012, ApJ, 757, 190

\bibitem[{{Deeming}(1975)}]{dem}
{Deeming}, T.~J. 1975, Ap\&SS, 36, 137

\bibitem[{{Demarque} {et~al.}(2008){Demarque}, {Guenther}, {Li}, {Mazumdar}, \&
  {Straka}}]{dem08}
{Demarque}, P., {Guenther}, D.~B., {Li}, L.~H., {Mazumdar}, A., \& {Straka},
  C.~W. 2008, ApSS, 316, 31

\bibitem[{{di Mauro} {et~al.}(2011){di Mauro}, {Cardini}, {Catanzaro},
  {Ventura}, {Barban}, {Bedding}, {Christensen-Dalsgaard}, {De Ridder},
  {Hekker}, {Huber}, {Kallinger}, {Miglio}, {Montalban}, {Mosser}, {Stello},
  {Uytterhoeven}, {Kinemuchi}, {Kjeldsen}, {Mullally}, \& {Still}}]{diMauro11}
{di Mauro}, M.~P., {Cardini}, D., {Catanzaro}, G., {et~al.} 2011, MNRAS, 415,
  3783

\bibitem[{{Dupret} {et~al.}(2009){Dupret}, {Belkacem}, {Samadi}, {Montalban},
  {Moreira}, {Miglio}, {Godart}, {Ventura}, {Ludwig}, {Grigahc{\`e}ne},
  {Goupil}, {Noels}, \& {Caffau}}]{dup09}
{Dupret}, M., {Belkacem}, K., {Samadi}, R., {et~al.} 2009, A\&A, 506, 57

\bibitem[{{Feroz} {et~al.}(2009){Feroz}, {Hobson}, \& {Bridges}}]{feroz09}
{Feroz}, F., {Hobson}, M.~P., \& {Bridges}, M. 2009, MNRAS, 398, 1601

\bibitem[{{Frohlich} {et~al.}(1997){Frohlich}, {Andersen}, {Appourchaux},
  {Berthomieu}, {Crommelynck}, {Domingo}, {Fichot}, {Finsterle}, {Gomez},
  {Gough}, {Jimenez}, {Leifsen}, {Lombaerts}, {Pap}, {Provost}, {Cortes},
  {Romero}, {Roth}, {Sekii}, {Telljohann}, {Toutain}, \& {Wehrli}}]{fro97}
{Frohlich}, C., {Andersen}, B.~N., {Appourchaux}, T., {et~al.} 1997, Solar
  Physics, 170, 1

\bibitem[{{Gabriel}(1994)}]{gabriel1994}
{Gabriel}, M. 1994, A\&A, 287, 685

\bibitem[{{Garc{\'{\i}}a} {et~al.}(2011){Garc{\'{\i}}a}, {Hekker}, {Stello},
  {Guti{\'e}rrez-Soto}, {Handberg}, {Huber}, {Karoff}, {Uytterhoeven},
  {Appourchaux}, {Chaplin}, {Elsworth}, {Mathur}, {Ballot},
  {Christensen-Dalsgaard}, {Gilliland}, {Houdek}, {Jenkins}, {Kjeldsen},
  {McCauliff}, {Metcalfe}, {Middour}, {Molenda-Zakowicz}, {Monteiro}, {Smith},
  \& {Thompson}}]{gar11}
{Garc{\'{\i}}a}, R.~A., {Hekker}, S., {Stello}, D., {et~al.} 2011, MNRAS, 414,
  L6

\bibitem[{{Garc{\i}a} {et~al.}(2014){Garc{\i}a}, {Mathur}, {Pires}, {Regulo},
  {Bellamy}, {Palle}, {Ballot}, {Barcelo Forteza}, {Beck}, {Bedding},
  {Ceillier}, {Roca Cortes}, {Salabert}, \& {Stello}}]{garcia14}
{Garc{\i}a}, R.~A., {Mathur}, S., {Pires}, S., {et~al.} 2014, ArXiv e-prints
  (1405.5374)

\bibitem[{{Garc{\'{\i}}a} {et~al.}(2014){Garc{\'{\i}}a}, {P{\'e}rez
  Hern{\'a}ndez}, {Benomar}, {Silva Aguirre}, {Ballot}, {Davies}, {Do{\u g}an},
  {Stello}, {Christensen-Dalsgaard}, {Houdek}, {Ligni{\`e}res}, {Mathur},
  {Takata}, {Ceillier}, {Chaplin}, {Mathis}, {Mosser}, {Ouazzani},
  {Pinsonneault}, {Reese}, {R{\'e}gulo}, {Salabert}, {Thompson}, {van Saders},
  {Neiner}, \& {De Ridder}}]{garcia13}
{Garc{\'{\i}}a}, R.~A., {P{\'e}rez Hern{\'a}ndez}, F., {Benomar}, O., {et~al.}
  2014, A\&A, 563, A84

\bibitem[{{Gilliland} {et~al.}(2010){Gilliland}, {Brown},
  {Christensen-Dalsgaard}, {Kjeldsen}, {Aerts}, {Appourchaux}, {Basu},
  {Bedding}, {Chaplin}, {Cunha}, {De Cat}, {De Ridder}, {Guzik}, {Handler},
  {Kawaler}, {Kiss}, {Kolenberg}, {Kurtz}, {Metcalfe}, {Monteiro}, {Szab{\'o}},
  {Arentoft}, {Balona}, {Debosscher}, {Elsworth}, {Quirion}, {Stello},
  {Su{\'a}rez}, {Borucki}, {Jenkins}, {Koch}, {Kondo}, {Latham}, {Rowe}, \&
  {Steffen}}]{gil10}
{Gilliland}, R.~L., {Brown}, T.~M., {Christensen-Dalsgaard}, J., {et~al.} 2010,
  PASP, 122, 131

\bibitem[{{Gruberbauer} {et~al.}(2009){Gruberbauer}, {Kallinger}, {Weiss}, \&
  {Guenther}}]{gru09}
{Gruberbauer}, M., {Kallinger}, T., {Weiss}, W.~W., \& {Guenther}, D.~B. 2009,
  A\&A, 506, 1043

\bibitem[{{Guenther}(1994)}]{gue94}
{Guenther}, D.~B. 1994, ApJ, 422, 400

\bibitem[{{Guenther} {et~al.}(1992){Guenther}, {Demarque}, {Kim}, \&
  {Pinsonneault}}]{gue92}
{Guenther}, D.~B., {Demarque}, P., {Kim}, Y.-C., \& {Pinsonneault}, M.~H. 1992,
  ApJ, 387, 372

\bibitem[{{Harvey}(1985)}]{har85}
{Harvey}, J. 1985, in ESA Special Publication, Vol. 235, Future Missions in
  Solar, Heliospheric \& Space Plasma Physics, ed. {E.~Rolfe \& B.~Battrick},
  199

\bibitem[{{Harvey} {et~al.}(1993){Harvey}, {Duvall}, {Jefferies}, \&
  {Pomerantz}}]{har93}
{Harvey}, J.~W., {Duvall}, Jr., T.~L., {Jefferies}, S.~M., \& {Pomerantz},
  M.~A. 1993, in Astronomical Society of the Pacific Conference Series,
  Vol.~42, GONG 1992. Seismic Investigation of the Sun and Stars, ed. T.~M.
  {Brown}, 111

\bibitem[{{Hekker} {et~al.}(2011){Hekker}, {Elsworth}, {De Ridder}, {Mosser},
  {Garc{\'{\i}}a}, {Kallinger}, {Mathur}, {Huber}, {Buzasi}, {Preston}, {Hale},
  {Ballot}, {Chaplin}, {R{\'e}gulo}, {Bedding}, {Stello}, {Borucki}, {Koch},
  {Jenkins}, {Allen}, {Gilliland}, {Kjeldsen}, \&
  {Christensen-Dalsgaard}}]{hek10b}
{Hekker}, S., {Elsworth}, Y., {De Ridder}, J., {et~al.} 2011, A\&A, 525, A131

\bibitem[{{Hekker} {et~al.}(2012){Hekker}, {Elsworth}, {Mosser}, {Kallinger},
  {Chaplin}, {De Ridder}, {Garc{\'{\i}}a}, {Stello}, {Clarke}, {Hall}, \&
  {Ibrahim}}]{hek12}
{Hekker}, S., {Elsworth}, Y., {Mosser}, B., {et~al.} 2012, A\&A, 544, A90

\bibitem[{{Herschel}(1801)}]{herschel}
{Herschel}, W. 1801, Royal Society of London Philosophical Transactions Series
  I, 91, 265

\bibitem[{{Houdek} {et~al.}(1999){Houdek}, {Balmforth},
  {Christensen-Dalsgaard}, \& {Gough}}]{houdek99}
{Houdek}, G., {Balmforth}, N.~J., {Christensen-Dalsgaard}, J., \& {Gough},
  D.~O. 1999, A\&A, 351, 582

\bibitem[{{Huber} {et~al.}(2011){Huber}, {Bedding}, {Stello}, {Hekker},
  {Mathur}, {Mosser}, {Verner}, {Bonanno}, {Buzasi}, {Campante}, {Elsworth},
  {Hale}, {Kallinger}, {Silva Aguirre}, {Chaplin}, {De Ridder},
  {Garc{\'{\i}}a}, {Appourchaux}, {Frandsen}, {Houdek}, {Molenda-{\.Z}akowicz},
  {Monteiro}, {Christensen-Dalsgaard}, {Gilliland}, {Kawaler}, {Kjeldsen},
  {Broomhall}, {Corsaro}, {Salabert}, {Sanderfer}, {Seader}, \&
  {Smith}}]{hub11}
{Huber}, D., {Bedding}, T.~R., {Stello}, D., {et~al.} 2011, ApJ, 743, 143

\bibitem[{{Huber} {et~al.}(2010){Huber}, {Bedding}, {Stello}, {Mosser},
  {Mathur}, {Kallinger}, {Hekker}, {Elsworth}, {Buzasi}, {De Ridder},
  {Gilliland}, {Kjeldsen}, {Chaplin}, {Garc{\'{\i}}a}, {Hale}, {Preston},
  {White}, {Borucki}, {Christensen-Dalsgaard}, {Clarke}, {Jenkins}, \&
  {Koch}}]{hub10}
{Huber}, D., {Bedding}, T.~R., {Stello}, D., {et~al.} 2010, ApJ, 723, 1607

\bibitem[{Jeffreys(1998)}]{jeffreys61}
Jeffreys, H. 1998, Theory of probability, Oxford Classic Texts in the Physical
  Sciences (New York: The Clarendon Press Oxford University Press), xii+459,
  reprint of the 1983 edition

\bibitem[{{Jenkins} {et~al.}(2010){Jenkins}, {Caldwell}, {Chandrasekaran},
  {Twicken}, {Bryson}, {Quintana}, {Clarke}, {Li}, {Allen}, {Tenenbaum}, {Wu},
  {Klaus}, {Van Cleve}, {Dotson}, {Haas}, {Gilliland}, {Koch}, \&
  {Borucki}}]{jen10}
{Jenkins}, J.~M., {Caldwell}, D.~A., {Chandrasekaran}, H., {et~al.} 2010, ApJL,
  713, L120

\bibitem[{{Kallinger} {et~al.}(2010{\natexlab{a}}){Kallinger}, {Gruberbauer},
  {Guenther}, {Fossati}, \& {Weiss}}]{kal10c}
{Kallinger}, T., {Gruberbauer}, M., {Guenther}, D.~B., {Fossati}, L., \&
  {Weiss}, W.~W. 2010{\natexlab{a}}, A\&A, 510, A106

\bibitem[{{Kallinger} {et~al.}(2012){Kallinger}, {Hekker}, {Mosser}, {De
  Ridder}, {Bedding}, {Elsworth}, {Gruberbauer}, {Guenther}, {Stello}, {Basu},
  {Garc{\'{\i}}a}, {Chaplin}, {Mullally}, {Still}, \& {Thompson}}]{kal12}
{Kallinger}, T., {Hekker}, S., {Mosser}, B., {et~al.} 2012, A\&A, 541, A51

\bibitem[{{Kallinger} \& {Matthews}(2010)}]{kal10d}
{Kallinger}, T. \& {Matthews}, J.~M. 2010, ApJL, 711, L35

\bibitem[{{Kallinger} {et~al.}(2010{\natexlab{b}}){Kallinger}, {Mosser},
  {Hekker}, {Huber}, {Stello}, {Mathur}, {Basu}, {Bedding}, {Chaplin}, {De
  Ridder}, {Elsworth}, {Frandsen}, {Garc{\'{\i}}a}, {Gruberbauer}, {Matthews},
  {Borucki}, {Bruntt}, {Christensen-Dalsgaard}, {Gilliland}, {Kjeldsen}, \&
  {Koch}}]{kal10b}
{Kallinger}, T., {Mosser}, B., {Hekker}, S., {et~al.} 2010{\natexlab{b}}, A\&A,
  522, A1

\bibitem[{{Kallinger} {et~al.}(2010{\natexlab{c}}){Kallinger}, {Weiss},
  {Barban}, {Baudin}, {Cameron}, {Carrier}, {De Ridder}, {Goupil},
  {Gruberbauer}, {Hatzes}, {Hekker}, {Samadi}, \& {Deleuil}}]{kal10a}
{Kallinger}, T., {Weiss}, W.~W., {Barban}, C., {et~al.} 2010{\natexlab{c}},
  A\&A, 509, A77

\bibitem[{{Karoff}(2012)}]{karoff12}
{Karoff}, C. 2012, MNRAS, 421, 3170

\bibitem[{{Karoff} {et~al.}(2013){Karoff}, {Campante}, {Ballot}, {Kallinger},
  {Gruberbauer}, {Garc{\'{\i}}a}, {Caldwell}, {Christiansen}, \&
  {Kinemuchi}}]{karoff13}
{Karoff}, C., {Campante}, T.~L., {Ballot}, J., {et~al.} 2013, ApJ, 767, 34

\bibitem[{{Kjeldsen} \& {Bedding}(1995)}]{kje95}
{Kjeldsen}, H. \& {Bedding}, T.~R. 1995, A\&A, 293, 87

\bibitem[{{Kjeldsen} \& {Bedding}(2011)}]{kjeldsen11}
{Kjeldsen}, H. \& {Bedding}, T.~R. 2011, A\&A, 529, L8

\bibitem[{{Kjeldsen} {et~al.}(2008){Kjeldsen}, {Bedding}, {Arentoft}, {Butler},
  {Dall}, {Karoff}, {Kiss}, {Tinney}, \& {Chaplin}}]{kjeldsen08}
{Kjeldsen}, H., {Bedding}, T.~R., {Arentoft}, T., {et~al.} 2008, ApJ, 682, 1370

\bibitem[{{Koch} {et~al.}(2010){Koch}, {Borucki}, {Basri}, {Batalha}, {Brown},
  {Caldwell}, {Christensen-Dalsgaard}, {Cochran}, {DeVore}, {Dunham},
  {Gautier}, {Geary}, {Gilliland}, {Gould}, {Jenkins}, {Kondo}, {Latham},
  {Lissauer}, {Marcy}, {Monet}, {Sasselov}, {Boss}, {Brownlee}, {Caldwell},
  {Dupree}, {Howell}, {Kjeldsen}, {Meibom}, {Morrison}, {Owen}, {Reitsema},
  {Tarter}, {Bryson}, {Dotson}, {Gazis}, {Haas}, {Kolodziejczak}, {Rowe}, {Van
  Cleve}, {Allen}, {Chandrasekaran}, {Clarke}, {Li}, {Quintana}, {Tenenbaum},
  {Twicken}, \& {Wu}}]{koch10}
{Koch}, D.~G., {Borucki}, W.~J., {Basri}, G., {et~al.} 2010, ApJ, 713, L79

\bibitem[{{Korzennik} {et~al.}(2012){Korzennik}, {Rabello-Soares}, {Schou}, \&
  {Larson}}]{Korzennik12}
{Korzennik}, S.~G., {Rabello-Soares}, M.~C., {Schou}, J., \& {Larson}, T.~P.
  2012, in Astronomical Society of the Pacific Conference Series, Vol. 462,
  Progress in Solar/Stellar Physics with Helio- and Asteroseismology, ed.
  H.~{Shibahashi}, M.~{Takata}, \& A.~E. {Lynas-Gray}, 189

\bibitem[{{Ludwig}(2006)}]{ludwig2006}
{Ludwig}, H.-G. 2006, A\&A, 445, 661

\bibitem[{{Ludwig} {et~al.}(2009){Ludwig}, {Samadi}, {Steffen}, {Appourchaux},
  {Baudin}, {Belkacem}, {Boumier}, {Goupil}, \& {Michel}}]{ludwig09}
{Ludwig}, H.-G., {Samadi}, R., {Steffen}, M., {et~al.} 2009, A\&A, 506, 167

\bibitem[{{Mathur} {et~al.}(2010){Mathur}, {Garc{\'{\i}}a}, {R{\'e}gulo},
  {Creevey}, {Ballot}, {Salabert}, {Arentoft}, {Quirion}, {Chaplin}, \&
  {Kjeldsen}}]{mat10}
{Mathur}, S., {Garc{\'{\i}}a}, R.~A., {R{\'e}gulo}, C., {et~al.} 2010, A\&A,
  511, A46

\bibitem[{{Mathur} {et~al.}(2011){Mathur}, {Hekker}, {Trampedach}, {Ballot},
  {Kallinger}, {Buzasi}, {Garc{\'{\i}}a}, {Huber}, {Jim{\'e}nez}, {Mosser},
  {Bedding}, {Elsworth}, {R{\'e}gulo}, {Stello}, {Chaplin}, {De Ridder},
  {Hale}, {Kinemuchi}, {Kjeldsen}, {Mullally}, \& {Thompson}}]{mat11}
{Mathur}, S., {Hekker}, S., {Trampedach}, R., {et~al.} 2011, ApJ, 741, 119

\bibitem[{{Michel} {et~al.}(2009){Michel}, {Samadi}, {Baudin}, {Barban},
  {Appourchaux}, \& {Auvergne}}]{mic09}
{Michel}, E., {Samadi}, R., {Baudin}, F., {et~al.} 2009, A\&A, 495, 979

\bibitem[{{Miglio} {et~al.}(2012){Miglio}, {Brogaard}, {Stello}, {Chaplin},
  {D'Antona}, {Montalb{\'a}n}, {Basu}, {Bressan}, {Grundahl}, {Pinsonneault},
  {Serenelli}, {Elsworth}, {Hekker}, {Kallinger}, {Mosser}, {Ventura},
  {Bonanno}, {Noels}, {Silva Aguirre}, {Szabo}, {Li}, {McCauliff}, {Middour},
  \& {Kjeldsen}}]{mig11}
{Miglio}, A., {Brogaard}, K., {Stello}, D., {et~al.} 2012, MNRAS, 419, 2077

\bibitem[{{Mosser} {et~al.}(2013{\natexlab{a}}){Mosser}, {Belkacem}, \&
  {Vrard}}]{mosser13}
{Mosser}, B., {Belkacem}, K., \& {Vrard}, M. 2013{\natexlab{a}}, in EAS
  Publications Series, Vol.~63, EAS Publications Series, 137--150

\bibitem[{{Mosser} {et~al.}(2013{\natexlab{b}}){Mosser}, {Dziembowski},
  {Belkacem}, {Goupil}, {Michel}, {Samadi}, {Soszy{\'n}ski}, {Vrard},
  {Elsworth}, {Hekker}, \& {Mathur}}]{mos13a}
{Mosser}, B., {Dziembowski}, W.~A., {Belkacem}, K., {et~al.}
  2013{\natexlab{b}}, A\&A, 559, A137

\bibitem[{{Mosser} {et~al.}(2012{\natexlab{a}}){Mosser}, {Elsworth}, {Hekker},
  {Huber}, {Kallinger}, {Mathur}, {Belkacem}, {Goupil}, {Samadi}, {Barban},
  {Bedding}, {Chaplin}, {Garcia}, {Stello}, {De Ridder}, {Middour}, {Morris},
  \& {Quintana}}]{mos12a}
{Mosser}, B., {Elsworth}, Y., {Hekker}, S., {et~al.} 2012{\natexlab{a}}, A\&A,
  537, A30

\bibitem[{{Mosser} {et~al.}(2012{\natexlab{b}}){Mosser}, {Goupil}, {Belkacem},
  {Marques}, {Beck}, {Bloemen}, {De Ridder}, {Barban}, {Deheuvels}, {Elsworth},
  {Hekker}, {Kallinger}, {Ouazzani}, {Pinsonneault}, {Samadi}, {Stello},
  {Garc{\'{\i}}a}, {Klaus}, {Li}, {Mathur}, \& {Morris}}]{mos12c}
{Mosser}, B., {Goupil}, M.~J., {Belkacem}, K., {et~al.} 2012{\natexlab{b}},
  A\&A, 548, A10

\bibitem[{{Mosser} {et~al.}(2012{\natexlab{c}}){Mosser}, {Goupil}, {Belkacem},
  {Michel}, {Stello}, {Marques}, {Elsworth}, {Barban}, {Beck}, {Bedding}, {De
  Ridder}, {Garc{\'{\i}}a}, {Hekker}, {Kallinger}, {Samadi}, {Stumpe},
  {Barclay}, \& {Burke}}]{mos12b}
{Mosser}, B., {Goupil}, M.~J., {Belkacem}, K., {et~al.} 2012{\natexlab{c}},
  A\&A, 540, A143

\bibitem[{{Muller}(1999)}]{muller99}
{Muller}, R. 1999, in Astrophysics and Space Science Library, Vol. 239, Motions
  in the Solar Atmosphere, ed. A.~{Hanslmeier} \& M.~{Messerotti}, 35--70

\bibitem[{{Pinsonneault} {et~al.}(2012){Pinsonneault}, {An},
  {Molenda-{\.Z}akowicz}, {Chaplin}, {Metcalfe}, \& {Bruntt}}]{pinsonneault12}
{Pinsonneault}, M.~H., {An}, D., {Molenda-{\.Z}akowicz}, J., {et~al.} 2012,
  ApJS, 199, 30

\bibitem[{{Samadi} {et~al.}(2013{\natexlab{a}}){Samadi}, {Belkacem}, \&
  {Ludwig}}]{sam13a}
{Samadi}, R., {Belkacem}, K., \& {Ludwig}, H.-G. 2013{\natexlab{a}}, A\&A, 559,
  A39

\bibitem[{{Samadi} {et~al.}(2013{\natexlab{b}}){Samadi}, {Belkacem}, {Ludwig},
  {Caffau}, {Campante}, {Davies}, {Kallinger}, {Lund}, {Mosser}, {Baglin},
  {Mathur}, \& {Garcia}}]{sam13}
{Samadi}, R., {Belkacem}, K., {Ludwig}, H.-G., {et~al.} 2013{\natexlab{b}},
  A\&A, 559, A40

\bibitem[{{Samadi} {et~al.}(2007){Samadi}, {Georgobiani}, {Trampedach},
  {Goupil}, {Stein}, \& {Nordlund}}]{sam07}
{Samadi}, R., {Georgobiani}, D., {Trampedach}, R., {et~al.} 2007, A\&A, 463,
  297

\bibitem[{{Smith} {et~al.}(2012){Smith}, {Stumpe}, {Van Cleve}, {Jenkins},
  {Barclay}, {Fanelli}, {Girouard}, {Kolodziejczak}, {McCauliff}, {Morris}, \&
  {Twicken}}]{smith12}
{Smith}, J.~C., {Stumpe}, M.~C., {Van Cleve}, J.~E., {et~al.} 2012, PASP, 124,
  1000

\bibitem[{{Stello} {et~al.}(2011){Stello}, {Huber}, {Kallinger}, {Basu},
  {Mosser}, {Hekker}, {Mathur}, {Garc{\'{\i}}a}, {Bedding}, {Kjeldsen},
  {Gilliland}, {Verner}, {Chaplin}, {Benomar}, {Meibom}, {Grundahl},
  {Elsworth}, {Molenda-{\.Z}akowicz}, {Szab{\'o}}, {Christensen-Dalsgaard},
  {Tenenbaum}, {Twicken}, \& {Uddin}}]{ste11}
{Stello}, D., {Huber}, D., {Kallinger}, T., {et~al.} 2011, ApJL, 737, L10

\bibitem[{{Thygesen} {et~al.}(2012){Thygesen}, {Frandsen}, {Bruntt},
  {Kallinger}, {Andersen}, {Elsworth}, {Hekker}, {Karoff}, {Stello},
  {Brogaard}, {Burke}, {Caldwell}, \& {Christiansen}}]{thy12}
{Thygesen}, A.~O., {Frandsen}, S., {Bruntt}, H., {et~al.} 2012, A\&A, 543, A160

\bibitem[{{Trampedach} {et~al.}(2013){Trampedach}, {Asplund}, {Collet},
  {Nordlund}, \& {Stein}}]{tra13}
{Trampedach}, R., {Asplund}, M., {Collet}, R., {Nordlund}, {\AA}., \& {Stein},
  R.~F. 2013, ApJ, 769, 18

\bibitem[{{V{\'a}zquez Rami{\'o}} {et~al.}(2005){V{\'a}zquez Rami{\'o}},
  {R{\'e}gulo}, \& {Roca Cort{\'e}s}}]{vaz05}
{V{\'a}zquez Rami{\'o}}, H., {R{\'e}gulo}, C., \& {Roca Cort{\'e}s}, T. 2005,
  A\&A, 443, L11

\end{thebibliography}

\end{document}